\pdfoutput=1
\documentclass[11pt]{article}

\usepackage[letterpaper,top=2cm,bottom=2cm,left=3cm,right=3cm,marginparwidth=1.75cm]{geometry}
\usepackage{xcolor}
\usepackage[colorlinks=true, allcolors=cyan]{hyperref}
\usepackage{xurl}

\usepackage[english]{babel}

\usepackage{amsmath}
\usepackage{graphicx}
\newcommand\comment[1]{\textcolor{red}{#1}}
\newcommand\marina[1]{\textcolor{blue}{Marina: #1}}
\newcommand\andrew[1]{\textcolor{olive}{Andrew: #1}}

\newcommand\george[1]{\textcolor{orange}{George: #1}}

\date{~}

\title{AI Horizon Scanning -- White Paper p3395\\ Part I.\ Areas of Attention}
\author{Marina Cort\^es,$^{1}$ Andrew R.\ Liddle,$^{1}$ 
Christos Emmanouilidis,$^2$\\
Anthony E.\ Kelly,$^3$
Ken Matusow,$^{1,4}$
Ragu Ragunathan,$^5$
Jayne M.\ Suess $^{5}$\\
George Tambouratzis,$^6$
Janusz Zalewski,$^{7,8}$
and David A.\ Bray$^{9,10}$
}

\begin{document}
\maketitle

\vspace*{-40pt}
\begin{center}
$^1$Institute for Astrophysics and Space Sciences, University of Lisbon, Portugal\\
$^2$ University of Groningen, The Netherlands\\
$^3$George Mason University, Fairfax, United States\\ 
$^4$
Synergycity, California, United States\\
$^5$Capitol Technology University, Laurel MD, 
United States\\
$^6$ Athena Research Centre, Athens, Greece\\
$^7$ Florida Gulf Coast University, United States,\\ 
$^{8}$ State Academy of Applied Sciences, Ciechanow, Poland\\
$^{9}$ Henry L. Stimson Center, Washington DC, United States\\
$^{10}$ National Academy of Public Administration, Washington, DC, United States
~\\
This version: \today
\end{center}


\begin{abstract}
Generative Artificial Intelligence (AI) models may carry societal transformation to an extent demanding a delicate balance between opportunity and risk. This manuscript is the first of a series of White Papers 
informing the development of IEEE-SA's p3995 \href{https://standards.ieee.org/ieee/3395/11378/}
{\it `Standard for the Implementation of Safeguards, Controls, and Preventive Techniques for Artificial Intelligence (AI) Models'} Chair: Marina Cort\^{e}s. In this first {\it horizon-scanning} we identify key attention areas for standards activities in AI. We examine different principles for regulatory efforts, and review notions of accountability, privacy, data rights and mis-use. As a safeguards standard we devote significant attention to 
the stability of global infrastructures and consider a possible over-dependence on cloud computing that may result from densely coupled AI components. We review the recent cascade-failure-like Crowdstrike event in July 2024,  
as an illustration of potential impacts on critical infrastructures from AI-induced incidents in the (near) future. 
Upcoming articles in this IEEE-SA White Paper series will focus on regulatory initiatives, technology evolution and the role of AI in specific domains.

\end{abstract}

\maketitle


\tableofcontents

\vskip 12pt

\subsubsection*{IEEE disclaimer}
This IEEE Standards Association (``IEEE-SA'') publication (``Work'') is {\it not} a consensus standard document. Specifically this document is NOT AN IEEE STANDARD.
This article solely represents the views of the set of authors within the IEEE p3395 Working Group, and does not necessarily represent a position of either IEEE or the IEEE Standards Association.
  Information contained in this ``Work'' has been created by, or obtained from, sources believed to be reliable, and reviewed by members of the activity that produced this ``Work''.
 Although the p3395 WG members who have created this Work believe that the information and guidance given in this ``Work'' serve as an enhancement to users, all persons must rely upon their own skill and judgment when making use of it. Please review Section~\ref{Sec:disclaimers} for an abridged version of the disclaimers in this ``Work''.  

\section{Introduction}
We live in an era of unprecedented technological leaps. As we move towards the end of this decade it will become increasingly challenging to discern whether online content is accurate or inauthentic \cite{panel4}. 
The massive increase in information we have access to has made understanding the world harder, not easier. Across higher education, culture, physics, to name a few areas, life has become very confusing \cite{YT:Weinstein}. The rapid development and adoption of Generative Artificial Intelligence (GenAI) technologies, currently based on Large Language Models (LLMs), adds both novelty and uncertainty to our interpretation of an already complex world. In the age of GenAI the distinction between human and bot-generated content, as well as authentic and inauthentic content is becoming increasingly challenging to societies across the board on the planet. 

The present authors are all ongoing volunteer contributors to the emerging IEEE\footnote{IEEE (The Institute of Electrical and Electronic Engineering) has over 400,000 members in 160 countries. Its Standards Association IEEE-SA issues global technical standards across all technological disciplines through voluntary membership participation.} standard p3395 {\it `Standard for the Implementation of Safeguards, Controls, and Preventive Techniques for Artificial Intelligence (AI) Models'}, Chair: Marina Cortês, \cite{P3395}. 
Ours was the first IEEE working group in AI standards, emerging at the time of the issuing of the US presidential Executive Order on AI development and use, on October $30^{th}$, 2023 \cite{RAGU2}.
While IEEE is a fully-international organisation independent of any national government, the Executive Order nevertheless provided important impetus and we held our kickoff meeting on November $2^{nd}$, 2023, just three days after its announcement.



Our aim here is to offer a cross-section of perspectives and realities, which reflect the diversity of nationalities and expertise within our working group (see also Section~\ref{Sec:Bios}). The reflections in this article are meant to be timely, but tentative, responses to a rapidly-changing landscape. The document reflects our optimism that our society is equipped to respond with the expertise necessary to address the organisational challenges posed by AI in its many forms.




\subsection{Our article series}

The present article focuses on highlighting a variety of attention areas that will be crucial to the beneficial development of AI technologies, and is the first in a sequence of works that explore different areas of AI. In companion articles we will provide: 
\begin{enumerate}
\item A description of issues around AI models from a regulatory perspective, as well as the national and international standards and initiatives to address them \cite{regulation}.
\item A description of technology developments in the global context \cite{techno-watch}.
\item A set of issues in the AI health domain, along with uses and associated risks \cite{AI-health}.
\item A detailed study on the impact on mental health of prolonged exposure to the online environment, with emphasis on the generative AI context, and a special focus on exposure to so-called AI-hallucinated content \cite{AI-mental-health}. 
\end{enumerate}

\subsection{Generative AI: Historical Context setting}
%


Large Language Models emerged from a long tradition of information theory and noisy channel models. Recursive neural networks used for machine learning-based natural language processing were augmented in 2017 with a mechanism that enabled highly-efficient parallel processing of input strings \cite{Vaswani}. Though the proposed `transformer' architecture was focused on text strings as inputs, the authors of Ref.~\cite{Vaswani} concluded the discussion of their approach with plans to apply their transformer architecture for inputs other than text. Indeed, in their NVIDIA-GTC plenary the seven Google-based authors noted that the architecture was designed to cover different modalities. The mathematical underpinning of the transformer-based technology is accessibly described in a \href{https://www.youtube.com/watch?v=aircAruvnKk}{series of YouTube videos} by 3Blue1Brown.

Based on transformers and pre-trained unsupervised deep-learning models, a class of AI technology, large generative AI models (LGAIM), has emerged with the capability to produce novel outputs of a sophistication not seen earlier \cite{Hacker}. The Large Language Models (LLMs) first came to wide public attention through generating output text from input text, particularly OpenAI's \href{https://chatgpt.com}{ChatGPT}. However many other types of input/output can be effectively treated as text (`tokenised') and processed the same way, giving M-LLMs (Multi-modal LLMs) which cover multiple modalities. A taxonomy of generative AI models released since 2021 identifies nine categories of modality \cite{GBGM}, though admittedly the final category is `other' so as to include those that do not fit the first eight. Those eight are text-to-image, text-to-three-dimensional views, image-to-text, text-to-video, text-to-audio, text-to-text, text-to-code, and text-to-science.

While the generative AI models present great potential to sectors of society such as education, research, and arts, they also pose significant risks \cite{Hacker,Suleyman,Bengio}. 
These include: 
\begin{itemize}
    \item generating erroneous answers and presenting fictitious text as real \cite{BaiDoo},
    \item outputting responses biased against gender, race, or ethnicity \cite{Kotek},
    \item unpredictable behaviour in complex and possibly critical environments, and
    \item  opening wide avenues for intentional misuse and exploitation.
\end{itemize}
Since generative AI models use publicly-available news articles, academic papers, social media posts, photos, and even chatbot chats as their training data, Lucchi \cite{Lucchi} argues that these tools pose legal issues concerning the ownership of the generated contents copyright, and the fair use of the training data. Regarding this last article we can view it in two ways: 
\begin{itemize}
    \item On one hand one can argue that the use is fair, since if the data are considered public domain then it can be used, provided sensitive personal information is removed; 
    \item On the other hand we can argue that the removal of sensitive information may depend on the privacy rights of individuals and the laws governing the country. For example, the EU's General Data Protection Regulation (GDPR) is far more restrictive than any US Federal Privacy Law. Please see our companion article \cite{regulation} for an extended discussion on international and national regulatory initiatives.
\end{itemize}
Ref.~\cite{Hayes} discusses the potential copyright infringements by generative AI when they use publicly-available content for training the models and suggests measures that are not burdensome on the development of AI. 

As additional technologies, such as generative AI, are added to the suite of AI capabilities, governments continue to track and provide guidance and mandates to benefit from the new opportunities and to manage the emerging risks better. For example, on October 30, 2023, the U.S. President issued an Executive Order titled ``Safe, Secure, and Trustworthy Development and Use of Artificial Intelligence'' \cite{RAGU2} that included several mandates to manage the risks of rapidly-growing generative AI models and to track and protect against the use of U.S.\ computing infrastructure for training AI models with malicious intent, such as cybercrimes.
There are two critical directives from this Executive Order. The first is that within 270 days of the order, the National Institute of Standards and Technology (NIST) under the U.S. Department of Commerce was to develop a risk management framework for generative AI models. This framework will be a companion resource to the prior framework covering the other AI models. The second one was that within 90 days of this executive order, the U.S. Department of Commerce secretary had to propose a regulation through which the U.S. providers of high-end computing infrastructure can 
be directed to report periodically on foreign clients building AI technologies for suspected malicious cyber activities. In July 2024 the \href{https://www.whitehouse.gov/briefing-room/statements-releases/2024/07/26/fact-sheet-biden-harris-administration-announces-new-ai-actions-and-receives-additional-major-voluntary-commitment-on-ai/}{White House} reported successful on-time completion of the required actions.
%
%





%
%
%
\subsection{The growing challenges of discerning authentic versus
inauthentic information and identity} 

It is increasingly apparent that we may be entering an extended era where inauthenticity and authenticity will be difficult to discriminate, this involving multiple forms of content including biometrics and more, which we will address in more detail in Ref.~\cite{AI-health}.

In isolated pockets, governments appear to be increasingly aware of this challenge. In pluralistic societies, such as the United States or the European Union, central governments traditionally have had a role to play in verifying the authentic vs.\ inauthentic nature of public information. However public trust in these centralised institutions as arbiters for identifying dubiously-sourced information content has been eroding.

To add to this challenge, both the political sphere and advertising market tend to benefit by presenting information as fully reliable when it can be argued that this is only somewhat the case (for example, when the distinction between fact and belief fades).  
One route to address this comes from autocratic governments, who have somewhat of a `home-field' advantage because they feature only one singular narrative. 
Autocratic regimes use specialised tools like filtering, censorship, and repression to ensure that only this narrative (authentic or not) is seen by a majority of their population. 

Pluralistic societies have a more difficult task ahead of them. It can be argued that the last ten years might pale in comparison to the challenges of fact-checking in a world flooded by both media and mediums of questionable authenticity.
In 2019--2020, the non-profit \href{https://peoplecentered.net}{People-Centered Internet coalition} proposed that a key software vendor, which owned a substantial volume of   \href{https://en.wikipedia.org/wiki/Customer_relationship_management}{Customer Relationship Management (CRM)} data including `out of band' questions as part of the CRM data package, could use `out of band' management questions with the aim of supplying an additional level of identity trust for the end entity of transactions. 
In the end that proposal was overshadowed by larger concerns that the software, given some of its features, could be misused in ways not intended by the software company. 

The extent and relative ease with which large volumes of public information available worldwide for citizens to see, hear, and sense, can be manipulated points towards difficult times ahead. 
Meanwhile, understanding the importance of triangulation in comparing information from sources to discern authenticity vs.\ inauthenticity remains time-consuming and hard. 
Who will help the public in a world experiencing a flood of questionable content, information, and identities? And who `watches' the adjudicators?

\subsection{Reports and perspectives on AI}

In the lead up to the horizon-scanning effort which is the focus of this and our later companion articles, we can highlight of a few developments as well as mentioning key milestones, such as global reports on the state of AI technology. In a companion article \cite{techno-watch}, we perform a more detailed overview of emerging technologies and industry trends.

In October 2023, the {\it `State of AI'} report \cite{stateofai} was published, a document of nearly of 200 pages, led by AI investors Nathan Benaich and the {\it Air Street Capital team}. The report mentions how computational technology might be, in current times, understood as the {\it ``new oil''}. It also emphasises how, generally speaking, actors in this highly-competitive industry are applying growing efforts to ``clone or surpass proprietary performance''. In particular the report highlights how  discussions around AI safety have 
{\it ``exploded into the mainstream, prompting action from governments and regulators around the world. However, this flurry of activity conceals profound divisions within the AI community and a lack of concrete progress towards global governance, as governments around the world pursue conflicting approaches.''} (extract from original in italic).

A more recent article (January 2024), ``Thousands of AI authors on the future of AI'' \cite{Graceetal}, provides a detailed assessment of the impact of AI technology via an opinion poll of AI researchers on a set of pre-defined questions, giving the time interval over which various AI milestones are expected to be achieved. 
The aggregate forecasts by the study give at least a 50\% chance of AI systems achieving several milestones by 2028, for example
\begin{itemize}
    \item autonomously constructing a payment processing site from scratch.
    \item autonomously downloading and fine-tuning a Large Language Model.
\end{itemize} 
Notably most milestones in Ref.~\cite{Graceetal} are predicted to be achieved significantly earlier than in a similar survey undertaken only one year before. While the questions asked might not be the most aligned with our purposes here, the discussion and statistical analysis is rigorous and meets peer-review standards.

A detailed and authoritative view on the current AI safety situation is provided by the May 2024 ``International Scientific Report on the Safety of Advanced AI (Interim report)'' led by AI pioneer Yoshua Bengio \cite{Bengio}. It focuses particularly on the technological status and outlook, and while highlighting the main areas of risk it largely restricts itself to possible technological, rather than sociological and regulatory, solutions.

An online `AI Risk Repository' has been developed at MIT \cite{RiskRep}. It classifies over 700 risks into a taxonomy featuring 43 categories within 7 broad domains:
\begin{itemize}
\item Discrimination and toxicity.
\item Privacy and security.
\item Misinformation.
\item Malicious actors and misuse.
\item Human--computer interaction.
\item Socioeconomic and environmental harms.
\item AI systems safety, failures, and limitations.
\end{itemize}
There are extensive links to existing literature and a suite of analysis tools.

\subsection{Research on artificial intelligence}

It is important to stress in this context that reliable and independent sources of data on state-of-the-art AI technology and its many manifestations are difficult to obtain. It is challenging to separate actual performance of current models amid hype on one hand, and doomerism on the other (see also Section \ref{Sec:AI-doom} for AI doomerism). For example, one leading organization providing objective data on disinformation, the Stanford Internet Observatory,  \href{https://www.platformer.news/stanford-internet-observatory-shutdown-stamos-diresta-sio/}{has recently been shut down} \cite{SIOshutdown}. Moreover, access to immense amounts of data on AI projects is restricted by business enterprises and is classified by those businesses as closed source and proprietary.

\subsection{Work of Service: the need for volunteer groups
}

In April 2024, the US National Academies of Science, of Engineering, and of Medicine organised a workshop entitled \textit{``Evolving Technological, Legal and Social Solutions to Counter Disinformation in Social Media --- A Workshop''} see Ref.~\cite{natacad}. This two-day workshop was designed to foster new research and collaborations, and build implementable solutions for a whole-of-society approach to mitigating disinformation and its detrimental effects. The webpage of the event can be found 
\href{https://www.nationalacademies.org/event/41384_04-2024_evolving-technological-legal-and-social-solutions-to-counter-disinformation-in-social-media-a-workshop}{here}, and all videos of the event are available \href{https://vimeo.com/showcase/11110544}{here}. 

This workshop is valuable because 
\begin{itemize}
    \item It was a global conference unifying several themes, industries, countries, and contexts.
    \item  The workshop's website includes a large number of citable sources and accompanying documentation.
    \item It offered rich content with speakers from different academic and non-academic fields, including various sectors in the industry, and governmental agencies.
    \item It supplied multi-national views from the US, EU, China, UK, and others.
\end{itemize}
Documentation on the findings of the workshop will be produced by the National Academies, in the form of Conference Proceedings. 

We will refer to the content of the workshop throughout these introductory remarks on our horizon scanning effort of the p3395 working group. In this section we wish to remark and echo a view which was shared there, and also in other volunteer standard groups (our own IEEE-SA activity, ISO, etc.). In their closing remarks both chairs of the workshop, Nobel laureate Saul Perlmutter (Lawrence Berkeley National Laboratory and UC Berkeley) and Joan Donovan (Boston University) have emphasised the need for funding and support for independent research on the effects and evolution of the technology. 

Eric Horvitz of Microsoft 
argued that this area of research is way underfunded. Multi-disciplinary meetings like Ref.~\cite{natacad} are needed every few months to do integrative work.
%
In Ref.~\cite{natacad}, Washington University's Kate Starbird highlighted the concern that independent researchers and horizon-scanning efforts are not incentivised to assess the evolving AI problem space nor to describe the risks of the technology in a way that avoids vested interests. She described that many working on these issues feel much like {\it ``walking along a precipice''}. There are those who would like to push independent researchers off the cliff, so-to-speak, so that the world continues to operate in a way that benefits them.

\section{Global algorithmic pipeline - critical infrastructures} 
\label{Sec:cascade}
Our planet is currently engaged in the development of a so-called  {\it `global operating system'}. That is to say, the setting up of networks of automated algorithmic pipelines reaching unprecedented scales of computerization and interconnectedness. It controls the movement of goods and services in both the real and the online environments, and relies on continuous flow of information amongst devices, servers, and satellites, all constrained by a complex network of international agreements and limitations.

Researchers and developers in the AI systems community are concerned about how long system coherence can be maintained across such a vast distributed systems network. We are currently facing unprecedented scales of automation. It is crucial to consider the various dimensions in which GenAI use in smart manufacturing contexts risks the disruption of global supply chains and provides challenges around security, quality and safety. 

\subsection{Cascade Failures}



In a worst-case scenario, algorithmic (or cascade) failure might lead to breakdown of supply chains. If such breakdown of supplies were to reach large enough scales, the structures that support societal organisation could  be at risk of rupture. Such scenarios were already discussed as possible causes of breakdown of supply chains by Tristan Harris and Aza Raskin on March 9th 2023, in \href{https://www.youtube.com/watch?v=xoVJKj8lcNQ&t=6s}{the AI dilemma} \cite{AIdilemma}.
The risk of system cascade-failures was pointed to in by Vint Cerf, widely celebrated as one of the Founding Fathers of the internet\footnote{Vint Cerf co-created, with Bob Kahn, the vital TCP/IP protocol that enables information to pass seamlessly around the internet. Their joint work has won numerous leading awards including the Turing Award and the Presidential Medal of Freedom.}, in a seminar in January 2024 Ref.~\cite{cerftalk}. Cerf alerted to the possibility of cascade failures affecting the coherence of global algorithmic pipelines. As an example Cerf mentioned the possibility of getting logged out of accounts by multi-factor identification (with online banking interfaces at highest risk). 
The  \href{https://www.bbc.com/news/world-us-canada-68672373}{crash of a container ship into a Baltimore bridge} in early 2024 caused some initial concern of the possibility of disrupted supply of goods.

Distributed systems are those most at risk, due to possible over reliance on cloud computing. Banking, telecommunication companies, airline and internet providers are all sectors requiring dedicated attention.
    The July 2024 global outage caused by a failed upgrade of the Crowdstrike Falcon security software, reported \href{https://www.bbc.com/news/articles/cpwdyxx0v64o}{here} and affecting almost 10 million computers, offering a taste of what might be ahead. 

Quoting from Chapter 16 of `Building Secure and Reliable Systems' \cite{BSRS},
``Each component plays a vital role in returning a disaster-stricken system to an operational state. Even if your Incident Response (IR) team is highly skilled, without procedures or automated systems, its ability to respond to a disaster will be inconsistent. If your technical procedures are documented but aren’t accessible or usable, they’ll likely never be implemented.'' Chapter 1 illustrates this with a hilarious, albeit consequentially minor, example of a historical cascade failure in Google's password management system.


The algorithmic closure of automated pipelines is therefore an essential priority, in our opinion. The large-scale automation of the pipeline of algorithms we are trying to implement at planetary scale must be ensured to connect end-to-end, and close consistently. This is even more true for algorithmic pipelines supporting `Critical Infrastructures'. These include infrastructures, systems and networks that provide essential services, regarded as essential for social and economic well-being and welfare. The \href{2024 CrowdStrike incident}{Crowdstrike incident}, caused by an error in a software update by the American cybersecurity company, is an example of one such accident that weak algorithmic closure might cause. Our goal, moving forward, is to support the lessening of the likelihood of system failures in large-scale automated systems, due to what we could call {\it algorithmic incompetence (ai)}. 
\subsection{Crowdstrike Falcon Security Update}
Before we go on to sketch our stance on AI safety, we review an IT system-update control failure event. We will consider the Crowdstrike event in July 2024, and discuss it in context of automated system updates, and feedback loops. 

Crowdstrike was not an implementation process that involved an AI component, per se. Rather it was a software evaluation issue, and 
further involved standard system privilege management activities. 
The incident affected 8.5 million bots across the Azure backbone worldwide, 
all requiring individual human intervention and performing a full system reboot, before functionality could be recovered. The effect was felt all the way across Azure, it comprising Azure's entire footprint. 
It is possible that Crowdstrike happened because we are in the era of massive automation. For our purposes here the question is not so much whether Crowdstrike is a direct failure from the tie-in to the AI component, but whether an incident of this scale is very likely to be uncorrelated from the current era we are living in, of large scale deployment of the novel generation of AI models.
Either way, we believe important lessons can be learned from the Crowdstrike incident:
\begin{itemize}
    \item explicitly forbidding single points of failure in critical infrastructures
    \item management of standard system-privilege activities
    \item software testing before roll out remains a priority, with automated system updates deemed the ones requiring the highest scrutiny.
\end{itemize}
Can an (over) dependence on AI components enhance the risks of this type of event happening in the near future? 
Crowdstrike may be an opportunity and a harbinger of a future where these sorts of instabilities in the pipeline and distributed systems and platforms become more commonplace, as well as harder for us to analyse and prevent. Here we draw attention to over-reliance on cloud computing, and implementation of robust protocols of automated system-update, particularly when infrastructures deemed critical to the normal functioning of societies are implicated, or at risk. 


\section{Overarching theme: who has regulatory sovereignty?}

We lay out the encompassing theme of our first article, in the regulatory domain, in Sections \ref{Sec:Government}, \ref{Sec:self-regulating-tech}, and \ref{Sec:Citizen-sovereignity}, and outline a broad-stroke division of organisational and regulatory initiatives.
Globally we observe an overarching dialogue between three paradigms for the choice of regulation level: state, market, and individual.
\begin{enumerate}
    \item Government-level sovereignty which we address in Section~\ref{Sec:Government}. We address topics of autocracy and mass scanning in Section~\ref{Sec:Autocracy}.
    \item Tech-level or platform-level sovereignty (Self-regulating industry, or private sector player) which we address in Section~\ref{Sec:self-regulating-tech}.
    \item Citizen-level sovereignty, namely the  accountability of user/producer in the AI value chain. We address this regulatory choice in Section \ref{Sec:Citizen-sovereignity}.
\end{enumerate}

There exist reports in the industry sector that regulatory efforts and policy towards AI might be rather premature, since the developers and platforms themselves describe or alert to their own lack of fully-consistent understanding of the upcoming dynamics of the technology as it knits together with communities across the globe.  

\subsection{Government-level sovereignty}\label{Sec:Government}
In government regulatory efforts one of the main challenges is to reduce the number of moving parts in the regulation pipeline and streamline the way the different parts interact. As in all top-down regulation schemes, one must not underestimate the non-deterministic nature of the systems that are being rolling out.

The good news is that we have democratised technologies that used to be available to a few. But inevitably we are going to have people that will use the technology as bad actors; how do we deter that, without losing the benefits? In the context of US regulation, it has been argued that one can wait for the full features of the European Digital Services Act (DSA, see Ref.~\cite{DSA}) to be fleshed out and use the best ingredients therein \cite{natacad}. Additionally, as a challenge to government-based sovereignty, it has been argued by tech-industry experts that
only the technically competent could regulate AI.

\subsubsection*{Which legislation would we choose to implement if the approval process was frictionless?}

In Ref.~\cite{panel4} 
the moderator Aziz Huq (University of Chicago Law) asked:
``What would be the top priority for passing legislation if we had friction-free regulatory approval? For example, which elements of European regulation would you propose to be adopted in the US?'' Meaning if our proposals were met with no resistance by the regulatory or legislative committees what would our proposals be? What concrete measures would be propose when political constraints are absent? The panelists answered in the following way:
\begin{itemize}
    \item David Bray advocated giving individuals  digital dignity, through the ability to control how their data can be used by others. Bray emphasised that our data is our voice, and as such, it is in our interest to work for our right to choose how our data is used.

    \item Jeff Kosseff 
    argued for the need to protect free speech as a fundamental requirement for democracy. Further the US Naval Academy scholar stressed the fact that we give up on democracy when we start to regulate free speech. We see this in the example of the Supreme Court in Brazil ordering certain accounts to be taken out of Twitter/X. In such examples the central authorities may issue orders for precision removal of specific content that does not favor those authorities.  He proposed a vast investment in media literacy education, for instance in schools and libraries.

    \item Josh Braun sought regulations of digital advertising space, limiting commercial data acquisition for targeting to rebalance contextual versus targetted advertising. The aim here is to make contextual advertising more attractive again. One goal is to help news outlets to be financed in such a way that does not depend on advertising,
    while also disfavouring hyper-targeting of advertising.
    He also picked up on the media literacy issue, noting by analogy with driving practices that while teaching safe driving (c.f.\ technology use) is a good directive, legislating for safer cars and roads needs also to be part of the solution.
    \item Nandini Jammi proposed the creation of a {\bf national registry of data brokers}, in order to make transparent which organisations are selling and trading our sensitive personal data, and to allow citizens to understand what data is being harvested. We need to be able to ascertain if our data is or is not being harvested, at any given time, in any given location.
    \item Nathalie Smuha stated that the US can learn from the European Union's data protection laws, particularly the GDPR (General Data Protection Regulation) rather than the DSA (Digital Services Act).
    Our personal data is our voice and is attached to our identity. In Europe one has the right to decide who has access to one's data and on what basis.
\end{itemize}


\subsubsection{Autocracy and mass scanning}
\label{Sec:Autocracy} 

In the context of AI regulation, countries without large AI sectors of their own (for example Chile and Brazil) are able to go in the direction of `autocratic' regulatory approaches in a way that the more AI-vested societies are not. The Chinese documentation on AI regulation is a good example of how an autocratic government can develop legislation packages, the broader context of the technology industry in China being described in, for example, Ref.~\cite{tencent}.
We discuss this in Ref.~\cite{regulation}. 
In November 2023 the EU Parliament's `civil liberties' committee rejected \href{https://edri.org/our-work/eu-parliament-committee-rejects-mass-scanning-of-private-and-encrypted-communications/}{mass scanning of private and encrypted communications} at government or parliament level. This is particularly relevant considering that in October 2023 there was ongoing tension in the EU on AI regulation at parliament level: the member states preferred tech-based regulatory efforts, while the EU parliament prefers citizen-based regulatory efforts, see Ref.~\cite{AI-bill-rights}.

\subsubsection{Central registry of users}\label{Sec:driverslicense}

Here we consider the ideas currently in circulation that specific technologies in the online environment would be deemed to require training and the obtaining of a {\it technology license} or {\it driver's license}. This training would educate the users of the technologies deemed as requiring such licensing and responsible conduct as attributed in Ref.~\cite{cerfMarch1st}. China \cite{china, CHINAreg} currently has the most advanced legislation in this area. In Ref.~\cite{regulation} we develop this nation-based legislative choice in more detail.

In Ref.~\cite{panel4} Jeff Kosseff argued that increasing media literacy to address misuses of the technology is not enough but is part of the problem. Nothing is going to fix misinformation but we can make progress. In his book {\it `Liar in a crowded theater'} \cite{Kosseff1} he defends that commercial speech receives protection. The US Congress passed laws during Covid to use the Federal Trade Commission to avert scams and give users the choice of what they want to see. Giving choice is not a legal solution, but it starts to address the problems. 
(see also Ref.~\cite{Kosseff2}).

One aspect of such training may be to avoid scams. We are all frequently confronted by scams and many people fall for them, the elderly being perceived as especially vulnerable. AI-fuelled scams, for instance using generated speech mimicking relatives, will be extremely challenging to recognise. The extent of scams is currently poorly researched and understood, despite initiatives such as the US-based \href{https://www.stopscamsalliance.org}{Stop Scams Alliance}. Learning to recognise and avoid conspiracy-based material is also vital. We need to develop the science of why people buy into misinformation. How can we aim to give the public autonomy to be freed from conspiratorial thinking?

On the topic of {\it `Technology licensing'} \cite{Bray-April11th}: there is a regime imposed by the Biden Executive Order which has legal force. Could the requirement for a driver's license be a way to amplify the reach of the US President's Executive Order? One could contemplate licensing data brokers. For example we require medical doctors to take a medical oath, and in that community there exist boards tasked with assessing whether a practitioner's professional oath has been broken. One could envisage a combination of the issue of certified licenses with the same inspiration as the oath taken by medical practitioners and the forming of the associated boards for assessing conformity to the rules in these licenses. 


\subsection{Tech-level sovereignty (Self-regulating industry) }\label{Sec:self-regulating-tech}

\subsubsection{Industry self risk-assessment}

In Ref.~\cite{natacad}, the founder of provenance and cryptographics technologies Eric Horvitz (Microsoft) alerted to the fact that steps to make progress in AI technology and AI regulation are being taken as if they are going to work, while in reality neither regulatory agencies nor technology companies have a complete view whether each of these actions will work. Horvitz defended the need for independent researchers to advise tech companies on what is the scientific process to move forward with caution; to develop experience and practices at scale to deal with large scale unknowns by new media and old media alike. How do we avoid entering a post-epistemic world that our grandkids will have to live in? Horvitz notes that AI systems can serve as potent weapons of persuasion and disinformation, with AI-generated content about people and events are already being employed in fraud, impersonation, and larger cyberinfluence programs.

Given all the ideas under discussion in the AI safety domain there is a large uncertainty on what it means to give adversarial systems (like the one authored by Horvitz himself \cite{HorvitzAdv,aaa,bbb}) the capability to explain and generate more believable stories, creating alternative histories and synthetic pasts.

In this variety of legislative effort 
there is an obligation on platforms to present risks in a way they know they can address. The counterpoint to address is the case where platforms inform only of the risks they can address and that do not threaten commercial interests. 
This means platforms would have to inform researchers which systemic risks they are potentially exposed to before they gain access to data. The call would be to open up platforms and their data to identify and mitigate new risks that arise.

\subsubsection{Open AI’s self risk-assessment}

As an example of industry self-monitoring of safety, OpenAI released a statement and article in January 2024 \cite{OpenAIbio}: ``We are building an early warning system for LLMs being capable of assisting in biological threat creation. Current models turn out to be, at most, mildly useful for this kind of misuse, and we will continue evolving our evaluation blueprint for the future.''.

Indeed none of the results show a statistically-significant effect in improving threat creation capability. However more important, and not commented on in their article, is that the sample is too small to be able to make any strong statement that there is {\it not} an effect. 
More generally, fear of finding an adverse effect might incentivize undertaking of self-assessment studies insensitive to exactly those effects.

A subsequent \href{https://www.washingtonpost.com/documents/2ea97cb4-34df-4bdd-a100-3572e93fdba1.pdf?itid=lk_inline_manual_4}{public letter to OpenAI} \cite{OpenAIsen} from several US Senators seeks clarification on their delivery against a wide set of promised AI safety goals.

\subsubsection{Microsoft's First Responsible AI Transparency Report}

On May 1st, 2024, Microsoft published their ``First Responsible AI Transparency Report'' \cite{MicrosoftAItransp}.
Eric Horvitz, as Microsoft Chief Scientific Officer, reported via $\mathbf{X}$, ``Today, we're publishing our first Responsible AI Transparency Report, offering an overview of our 8 year journey with defining our AI principles, formulating our programs and processes, and learning and evolving over time.''

\subsubsection{Technology default settings}


In Ref.~\cite{natacad} Nandini Jammi
describes  a tech-company attempt to lock consumers into monthly payments, without them being aware. This can be achieved by various means, for example the company can specify the default setting of the application as {\it `monthly subscription'}, while not denoting this setting explicitly in the interface with the consumer. Google responded quickly to halt this action. There are ways to holding the industry and companies therein accountable. One such example is the {\bf issuing of certifications on brand safety}.

\subsubsection{Post-nation states and single-actor economies}

Of the three paradigms listed here we are excluding debates involving so-called post-nation states. The new forms of generative AI (GenAI) have broken down the boundaries that once separated international and intranational communities, see for example Ref.~\cite{Bray-April11th}. Individual actors may possess considerable control over international access to technologies, for example the Starlink satellite internet system \cite{Starlink}. 

\subsection{Citizen-level sovereignty}\label{Sec:Citizen-sovereignity}



Citizen-level sovereignty arises through giving citizens the opportunity to take responsible action and choose accountability for their choices working together towards the alignment with overall social benefit. 
In that context we can include the same personal accountability in online environment that has been known to be part of social fabric, as argued by UC Berkeley's Brandie Nonnecke in Ref.~\cite{natacad}. Items discussed in the context of individual accountability include the assignment of credit to technology users that identify themselves strongly in online interactions, as for example proposed in Ref.~\cite{cerftalk}.

In the past we have seen that those with the means to record history also have the means to narrate it in their favour. 
In our society we are witnessing widespread lack of trust in institutions. For instance, there are specific components of \href{https://en.wikipedia.org/wiki/Section_230}{section 230} (part of the US  Communications Decency Act of 1996) that incentivise individuals to discern which sets of assumptions they wish to live under. 
In the scope of citizen accountability there is the need for studies on what happens when we give individuals control over their offered content. 

Lastly, in a more generic viewpoint expressed in 
Ref.~\cite{cerftalk}, Vint Cerf argues that regulation should focus on the use of the technology rather than on the technology itself, which we may not be able to understand. 

We also remind the reader that in acompanion article \cite{regulation} we explore national and international standards and regulatory initiatives in further detail.
%
%
%


\section{Accountability, anonymity, and privacy}

To what extent can entities, individual or corporate, be held responsible not only for their own actions, but the actions of their various AI agents? To what extent, if any, is such accountability consistent with anonymity of action? And if responsibility/accountability demands there is no anonymity, what are the implications for privacy and our rights over our own data? All these issues are heightened by the ever-evolving AI landscape.

\subsection{Accountability}
\label{Sec:accountability}

In the tech industry context accountability is something of a trigger word, due to its intimate connection to anonymity. The latter raises the values associated to freedom of speech and the existence of whistle-blowers, which are amongst the key tenets in pluralistic societies. 

A self-consistent attribution of accountability of individual citizens (users/producers) requires the formulation of metrics for individual/collective accountability of user/producer in the AI value chain. Such an algorithm for assignment of accountability is not yet available, nor is its timescale or even feasibility readily evaluated. 

Accountability is often mentioned in connection to a central registry of users, see e.g.\ Ref.~\cite{CHINAreg}, akin to a driver’s license (see also Section~\ref{Sec:driverslicense}). A new legislation package by the Chinese government is apparently leading the initiative with the (controversial) central registry. In our companion article \cite{regulation} we discuss this legislation in more detail. 

When assigning accountability of user/producer in the AI value chain, we can choose citizen-based agency: giving agency to individual users, or self-regulating users. We discussed this in Section~\ref{Sec:Citizen-sovereignity}.

\subsection{Anonymity and freedom of speech}
\label{Sec:Anonymity}

On the opposite side are arguments that we also give up anonymity when we operate a car that can be tracked by its license plate, see Section~\ref{Sec:accountability}. Also regulation of speech might be used in autocratic forms.

Jeff Kosseff (US Navy) released a book in late 2023 entitled {\it `Liar in a Crowded Theater: Freedom of Speech in a World of Misinformation'} \cite{Kosseff1}. In Ref.~\cite{natacad} he mentions that Section 230 (in the US context) addresses the issue of liability, and further that freedom of speech is constitutionally protected.

In Ref.~\cite{natacad} Aziz Huq states the US body defining what counts as free speech is the Supreme Court. Also important to take into account is the fact that free speech is not just a simple concept, but rather an evolving one. Nandini Jammi, founder of \href{https://checkmyads.org}{{\it CheckMyAds}} and who works on the campaign \href{https://x.com/slpng_giants}{{\it `Sleeping giants'}} that advocates for accountability of media, social, and advertising platforms, also argues strongly for Free speech in Ref.~\cite{natacad}.

In Ref.~\cite{cerftalk}
Vint Cerf was interviewed by the Burnes Center for Social Change in a seminar entitled {\it ``The Internet We Deserve: How the History of the Internet Could Inform the Future of Democracy and AI''}, covering a variety of topics. Cerf briefly mentions how a possibility for addressing the spread of deepfakes is the favoring of users who identify themselves strongly online. Cerf describes how this option sacrifices the right to anonymity to a certain extent, with possible detriment towards whistle-blowers, and more.

\subsection{Privacy -- Data subject rights}\label{Sec:Privacy} 

It is becoming increasingly difficult to believe that none of our data is being captured, at any given time, by whichever device and platform we are using. This is especially relevant in the health domain.
Privacy regulations vary amongst countries. An example is all EU countries are governed by the General Data Protection Regulation (GDPR), whereas the US does not have a federal regulation protecting citizens' privacy. In the US there are soft federal regulations, though each individual state has the power to enact privacy regulations (e.g.\ California enacted a law similar to GDPR).

In Ref.~\cite{natacad} Joan Donovan raised a question about biometric informational privacy and whether that aspect of privacy should be a standard part of the privacy-related package of data legislation. Donovan cites Kashmir Hill's book \cite{KHillbook} on Clearview AI's face-recognition capabilities as an example. China in particular is recognised as having the best-developed AI-powered facial recognition and tracking capabilities, which are being widely deployed.

%

In Ref.~\cite{BrayDignity},  David A. Bray summarises a People-Centered Internet conversation about the recent Executive Order on AI Safety, as well as what questions remain and the need for additional activities around data dignity and more. In the space of privacy and data subject rights there is the effect of forces constantly leveraging against each other. The reality persists that consumers do not understand how their personal data is being utilised in the advertising space generally, and much more so in the context of generative AI.

\subsection{Responsibility}

In Ref.~\cite{natacad} Susan Silbey (MIT) argued that there exists no system or algorithm for attributing responsibility in the context of harmful consequences, resulting from actions by a user/producer of the technology (whether intended or unintended). Silbey argues that AI technology is currently disrupting the normal patterns of human life. Furthermore, it is ignoring the normal patterns of social life, and it is doing so based on the holding of anonymity. 

The reason anonymity (see also Section~\ref{Sec:Anonymity}) is such a fragile subject is because we witness societies all over the world who have given up anonymity, only to be followed by the government abusing their power. We have plans for how to deal with the technology and how to curb the challenges accompanying it, but few of these schemes include the accountability factor. 
The problem space is continuously evolving even as we write this section of the White Paper. However the technology is evolving largely protagonist-free, making it easy to operate anonymously. All societies that have evolved and thrived until today are based on the foundational concepts of  accountability and trust. This threat is evolving freely since we have not developed a measure for attributing responsibility to multiple actors in the AI product pipeline. The social sciences have historically developed a multiplicity of tools to create responsibility, and address systemic consequences, see for example
1936 American Sociological Review, {\it ``There are always unanticipated consequences of social action''} \cite{Merton}.


\subsection{Transparency}
\label{Sec:Transparency}

Transparency reports need not be be deemed as system vulnerabilities, nor a threat to the business model of commercial companies, .
Facebook demonstrated this in 2009 with Facebook \href{https://facebook.github.io/prophet/}{Prophet}, which was open sourced from 2017 onwards (though Meta announced in 2023 that they plan no further development of the tool). Transparency reports need to be given to those in charge of the model and to the regulators,
possibly also in the context of open-source models. 

In new developments the major technology companies driving AI have, as of July 2024, initiated a joint open source group on AI security: \href{https://campustechnology.com/Articles/2024/07/19/Tech-Giants-Form-Open-Source-AI-Security-Group.aspx}{Coalition for Secure AI (CoSAI)}. This initiative is primarily aimed at helping the developers build their AI systems through standardisation and shared knowledge of risk. From a user perspective it may be most notable for the following list of items it does {\it not} include: ``The project does not envision the following topics as being in scope: misinformation, hallucinations, hateful or abusive content, bias, malware generation, phishing content generation or other topics in the domain of content safety.''~\cite{CoSAI}.

\subsection{Open source}\label{Sec:open-source}
 
While some AI models are classified as `open source', models released by some businesses may be better classified as `open license', since the for-profit entities often place restrictions on their use and applications. Further, the benchmarks for ranking publicly-available AI models have proven controversial and are under revision (see for example the Hugging Face LLM leaderboard blog \cite{huggingface}.). 

Therefore, since the data sources available to the authors on AI models'  current and putative future states are contested, discussions around AI transparency, capabilities, and accountability must remain tentative. 

In short, in the field of AI research, it is challenging to separate valid information from misinformation, disinformation, or malinformation.  Malinformation is information that is understood to be factual, but is either oversimplified, or access to it is controlled with self-serving intent as is described \href{https://reason.com/2023/03/22/the-crusade-against-malinformation-explicitly-targets-inconvenient-truths/}{in this article} in the journal \href{https://reason.com}{{\it `Reason'}}. In the next Section~\ref{Sec:misinformation} we begin to address each of these terminologies associated with use of AI data.

\section{AI misuse, mal-use, disuse; misinformation; data poisoning}\label{Sec:misinformation}


In 
Ref.~\cite{cerftalk},
Vint Cerf argues that these are exciting times for new machine-learning algorithms but they also pose risk to democracy due to misinformation, disinformation, and fragmentation of data. There is a general lack of understanding of the dynamics of the technology to inform legislation and law. 
%
%
\subsection{Misinformation, disinformation} \label{Sec:misinformation2}

In July 2023 the Harvard Kennedy School Misinformation Review published ``A survey of expert views on misinformation: Definitions, determinants, solutions, and future of the field'', \cite{harvard-misinformation}. We will focus on this publication as portray the challenges facing the field of misinformation studies. The favoured definition in the article was `False and misleading information'. However this appears to rely on the additional assumption that the misinformation is spread unknowingly, without intent to mislead. Therefore we assume that if misinformation is disseminated with intend to mislead, it becomes {\it `disinformation'}.
The quality of the study notwithstanding, it gives us an opportunity to raise a number of  questions (on the above survey or others in this field of study)
\begin{itemize}
    \item {\bf Misinformation Experts}\\
    Who are the ``misinformation experts'' aligned with, and which external factors?
    \item {\bf The Appropriately Informed}\\
    Could the sample be said to, supposedly, constitute or represent what we could call ``The Appropriately Informed''?
    \item {\bf Critical reflection}\\ 
    Did the study use multiple choice questions or actual cases to tease out nuance and expose the logic of the aligners?
    \item {\bf Gender}\\ 
    ``We did not consider gender in the recruitment of participants and do not have access to the gender breakdown of the respondents we contacted— it is thus unclear whether the gender imbalance in the survey is an artifact of our recruitment techniques or not.''
    \item {\bf Political affiliation, and North and Global South}\\
    ``The sample of participants covers a large number of countries with a bias towards Western liberal democratic countries. 
    \begin{itemize}
        \item Experts were from the United States (43), United Kingdom (16), France (14), Germany (13), Canada (5), Australia (5), Italy (5), Brazil (4), Netherlands (4), Israel (4), Spain (3), Switzerland (3), Austria (3), China (2), India (2), Singapore (2), Chile (2), Argentina (1), Finland (1), Hungary (1), Iran (1), Ireland (1), Lebanon (1), Norway (1), Pakistan (1), Russia (1), Turkey (1).
        \item Experts leaned strongly toward the left of the political spectrum: very right-wing (0), fairly right-wing (0), slightly right-of-center (7), center (15), slightly left-of-center (43), fairly left-wing (62), very left-wing (21).''
    \end{itemize}
\end{itemize}
Critical self-reflection by the researchers on issues like the set above may be very useful for disentangling inherent or implicit (internal) biases.
In Ref.~\cite{natacad},  Nandini Jammi 
describes the priorities of de-monitizing disinformation and disempowering actors promoting election misinformation. We need to engage public collectives to address the challenges of how misinformation spreads in evolving backgrounds and context-dependent interpretations.

Another misinformation-related problem concerns the propensity of LLMs to generate erroneous or misleading content (hallucinations) which may even be exploited for malicious applications (e.g.\ Ref.~\cite{Augenstein}, which focuses on the risks of LLMs for inaccurate, misleading, or entirely fabricated content). As noted in Ref.~\cite{Augenstein}, ``Search engines such as Google and Bing have historically been seen as reliable gateways to authoritative information sources; while in certain cases they have helped amplify disinformation. They are more reliable as they give a clear indication of their sources, unlike ChatGPT. However, as LLM-based chatbots become increasingly used for information-seeking, there is a risk that the public may receive unreliable information through a modality that has traditionally been trustworthy''. Also cross-lingual queries of ChatGPT have resulted in completely different (and contradictory) responses in different languages \cite{GeorgeRef1,GeorgeRef2} and a substantial drift of responses for a given language over a period of six months. 


We believe that, first and foremost an assembly of members working together towards creating a global standard must have a self-examination technique that produces a body of assumptions (``the {\it prior}'') that are put forward before the results are presented. The results of the standard and of the working group are then to be interpreted in the knowledge of the {\it prior} assumptions, belief, or context under which those results are valid or proposed. 

\subsection{Disinformation}

In Ref.~\cite{natacad} Eric Horvitz described 
how, at the World Economic Forum in 2019, he was the first person to show Tony Hall --- then Director General of the BBC --- a deepfake. 
As a news broadcast lead, the Director General was worried to realize that in the coming future one would possibly not be able to trust most information sources. The adequate strategy envisioned would be to have {\it `boots on the ground'} for personal verification of every news piece.\footnote{\href{https://vimeo.com/showcase/11110544/video/936224823/embed}{Video here. 
}}
At that time Hall became a collaborator of cross-organisational effort to build media provenance technologies --- see also Section~\ref{Sec:Provenance} where we discuss provenance in more detail. 
Horvitz argues that to address the challenges of misinformation and disinformation one can deploy a variety of approaches: policies to fund media literacy; alliances formed to detect and share, and to shut down obvious flows of obvious disinformation; content moderation; education intervention; technological intervention.


In Ref.~\cite{natacad} Bray argues that {\it ``The good news is we have democratized technologies that used to only be available to exquisite, nation-state capabilities, such as the CIA and KGB.\footnote{Discussion panel: \href{https://vimeo.com/showcase/11110544/video/936226597}{\it `Regulatory and Other Incentives and Disincentives for Behavior Change',7mins:30secs.}} The bad news is we have democratized all those capabilities, but we have not democratised information discernment: how to make sense of all these different feeds including intentional and accidental disinformation.''}

Bray focuses on bad actors that may use disinformation to disrupt structures at national defense level, at law-enforcement level, or simply to erode civil norms. Bad actors may come from domestic sources, non-state, international or network sources, or foreign-state sources. To address such challenges we must devise ways to involve citizens at the individual level to work collaboratively on solutions. In this respect the assessment of the quality of information is argued to be a task to be taken at the citizen-level, therefore endorsing approaches as in Section~\ref{Sec:Citizen-sovereignity}.

Such an endeavour is challenging or even ineffectual if enforced at government level (see Section~\ref{Sec:Government}), because it would be perceived as surveillance and information control. What is worse, it could actually {\it become} surveillance and information control, in the support of mass-surveillance cultures, see Section~\ref{Sec:Autocracy}. 

In this regard one cannot rely on platform control either (see Section~\ref{Sec:self-regulating-tech}) because when platforms report to shareholders there may be a case of misaligned incentives.
In Bray's opinion, involving the public is the way forward, by operationalizing non-profits and laboratories to devise means of enhancing digital dignity of individual citizens and involving them in the process. 
Potentially crowd-sourcing the spotting and tipping off of these activities can be encouraged, so that users can report if they see some potential misinformation content. We also might employ professional certified data scientists to undertake this effort. 

Ultimately we must learn the lessons of the last decade: if your data is taken away from you in a way that has not involved you as a stakeholder, then your voice has been taken away.

\subsection{Public assessment of information quality}

Adoption of information-quality standards is often argued to represent a compelling way for the technology sector to address the increasing challenges facing individual citizens in discerning between credible and non-credible, non-authentic information. We must find ways to make citizens better evaluators of information quality, as argued by Bray and Cerf in {\it Tech Should Advance Standards to Assess Information Quality} \cite{braycerfmisinformation}.

Fighting misinformation requires transparency (see also Section~\ref{Sec:Provenance}), standards, and empowering individuals with methods and tools for critical analysis by themselves. What the strategies would be for such a goal to be attained in useful time is at this time unclear to us.

In a fruitful discussion in Ref.~\cite{natacad} Joan Donovan (Harvard’s Shorenstein Centre on Media, Politics and Public Policy) debated questions like:\footnote{\href{https://vimeo.com/showcase/11110544/video/936224823/embed}{\it `Looking Forward: Reflections from the Workshop Planning Committee'.}}
\begin{itemize}
    \item Do citizens have the right to truth? \item Is the responsibility of the information conduits, social media and search engines, to put more work onto the ontologies and ranking systems that support fact-based content? 
    \item Can information affect the result of an election, as Samuel Woolley and collaborators pointed out 8 years ago \cite{woolley}? 
    \item What are the epistemologies and methodologies that we as societies possess for arriving at the truth? 
    \item Do we have a clear understanding as societies, or discussion spaces in the lack thereof, to debate on the nature of truth? 
    \item Do we agree that citizens have a so-called {\it `right to be dis-informed'}?
\end{itemize}
Donovan argues that we should bear in mind the importance to keep the human right to be dis-informed, and the right to kinds of information other than the {\it status quo} version. 
At the same time, we must bear in mind that over-regulation of information on the internet may reduce it to an entirely bureaucratic shopping mall. It is important to keep has some room for excitement and entertainment. A prevalent frustration amongst media and public policy theorists is that we did not yet crack the art of providing the public with T.A.L.K.: ``timely accurate local knowledge'' \cite{Donovan1,Donovan2}. Audiences at large need this in order to make informed democratic decisions.

As a final remark in Ref.~\cite{natacad} Donovan asks the core question {\it ``What are the durable, truth-telling institutions that we are going to need to rely on as we think about the supply side and the demand side of misinformation?''}.

\subsection{Technology misuse and bad actors}
%
In Ref.~\cite{Bray-April11th}, David Bray notes the increasing volume and sophistication of cyberattacks and particularly ransomware attacks, affecting both corporations and public/governmental entities. The cost of ransomware attacks is doubling annually. He also notes that Generative AI is open to exploitation by authoritarian regimes, for example surveillance of dissenting citizens.

Election interference is another topical issue, both from competing factions within countries and from other states seeking either direct advantage from an influenced outcome or simply to destabilize. Deepfakes, for example assessed by Moody's in a report discussed \href{https://www.cnbc.com/2024/07/10/election-deepfakes-undermine-institutional-credibility-moodys.html}{here} by CNBC, can at the very least undermine the confidence in election outcomes, while potentially altering the outcome of close contests. A widely-reported deepfake robocall simulating the voice of then-candidate Joe Biden subsequently led to the \href{https://www.nbcnews.com/politics/politics-news/steve-kramer-admitted-deepfaking-bidens-voice-new-hampshire-primary-rcna153626}{levying of a six million dollar fine} by the Federal Communications Commission.


\subsection{Data poisoning: Attacks on LLMs}\label{Sec:poison}

Nefarious use of AI technology could result in the use of poisonous data (ill-advised, specifically-formatted datasets) used to train an LLM with intent of modifying the model's behaviour. 
The concept of LLM data poisoning was studied and forensic restoration was addressed by Shan et al.~\cite{Chan-LLMAttacks}. The paper emphasized forensics tools to detect and restore from data poisoning attacks. As LLMs are `trained', there are both structured and unstructured methods used. When using structured datasets, there could be nefarious or erroneous data ingested to the model, causing incorrect or unexpected outputs. The study performed identified several methods using an iterative, clustering and pruning of data samples to restore models to pre-poisoned state at a rate of 98 percent.

The methods presented were focused on restoration; however, the more desirable controls would protect the servers hosting the LLMs using strict access controls, hardware hardening, intrusion prevention, intrusion detection, and activity monitoring controls. Additionally, there was not a discussion of the intent (nefarious) or lack of intent (potentially incompetence) resulting in the `poisoned' model. 

\subsection{Persuasive technology}
\label{s:persuasive}

It is quite plausible that AI systems will develop considerable persuasive capabilities \cite{persuade1}, though evidence that existing systems have them is weak at best \cite{persuade2}. Yuval Noah Harari has expressed this fear \href{https://www.youtube.com/watch?v=x6tMLAjPVyo}{in a YouTube interview}: ``{\it We are now hackable animals, that we have the technology to decipher what you think, what you want, to predict human choices, to manipulate human desires, in ways that were never possible before.}''  

In any case, we always need to be aware that whenever any service is provided to us apparently for free, it is because the providers believe that with the data they extract from us, they can change our behaviour in a way that will generate net income to them. Jaron Lanier in particular has been very clear on this (see his contributions in the film \href{https://www.thesocialdilemma.com}{The Social Dilemma}): it is not your data alone that is of value to the platform, it is the fact that they are able to use it to change your behaviour in a way that is ultimately profitable for them and their paying (e.g.\ advertising) clients. 

Major internet services have existed on this basis for decades already. How much more effective AI will make this type of business strategy is something we can only guess.

\section{Output provenance: addressing AI bias, traceability, and transparency}\label{Sec:Provenance}

The obtaining of data provenance from a given output allows for program errors including bias to be corrected. Provenance research is an effort into explainability of model output, making a connection between training dataset and model output. 

In Ref.~\cite{natacad} Microsoft lead Eric Horvitz, one of the founders of the cryptographic digital provenance technologies, appeals for support for provenance research. He argues\footnote{\href{https://vimeo.com/showcase/11110544/video/936224823/embed}{\it `Looking Forward: Reflections from the Workshop Planning Committee.}}
that regulatory initiatives like the US Executive Order are spelling out directives for action in the generative AI context, but an important argument to consider is that industry leaders and developers cannot anticipate at this point how those actions might play out in the long run.

In September 2023, Microsoft's Jaron Lanier delivered a UC Berkeley seminar entitled {\it Data Dignity and the Inversion of AI} \cite{JaronUCB}, in which he made the following points:
\begin{itemize}
\item In regions of parameter space where the dataset is sparse, there is not much antecedent data, and output might be abnormal. There is an opportunity to create a commercial opening here, by calling for data of the kind that the model lacks. 
In the present context there is no way to do this. We have a sense of a black-box mechanism that produces outputs, and we don’t know how it does it, contributing towards a mystic vision of generative AI that {\bf works against understanding of the models being used by the public at large}.

\item We can work towards considering LLMs as a social collaboration, instead of a mysterious creature. This can be done by inverting the way we look at the models.

\item How do we address the fact that training datasets are predominantly white and male and western? By making the training data explicated! When we get an output of the system, one should be able to get a characterization of the key antecedent examples that influenced the result.
   
\item The problem when we give the impression that GPT is a mysterious, infinitely-large, oracle that has a trackless interior that no one can interpret is that when you then complain about bias, your only resort is to try to slap another AI on the output to try to catch the bias, which leads you back into the genie problem.
Why can’t we be motivated to make the training data work better for society? 
\end{itemize}

In his UC Berkeley talk, Lanier, like Horvitz, stresses how the tech industry is in need of researchers for conducting output provenance studies. In particular the technical feasibility of tracing output at various levels needs to be explored, and made viable, with Microsoft willing to fund research by independent researchers. 


In his January 2024 interview Vint Cerf \cite{cerftalk} also focused on the relevance of obtaining provenance of model outputs. What information contributed to a given output, and what was its source and the way in which it was corroborated, all contribute towards ensuring trustworthiness, reliability, verifiability, and the validity of information produced by LLMs. 

The challenges on provenance research may stem from the possibility that output provenance is only possible if there is a decision tree underneath. In probabilistic paradigms this is very difficult. As far back as 10 years ago it was very already difficult to attribute output provenance to models that had only six hidden layers. 

\section{Academia}

\subsection{Journals}

Academia is an area experiencing significant threats from the ease with which plausible but fake content can now be generated. Journals are experiencing significant volumes of fraudulently created articles, in some cases ultimately leading to the shut-down of entire journals. As an example, \href{https://www.wsj.com/science/academic-studies-research-paper-mills-journals-publishing-f5a3d4bc}{Wiley} is said to have retracted more than 10,000 articles in the last two years and closed 19 journals, while Institute of Physics Publishing discovered nearly 900 fraudulent papers in 2022, a spokesperson commenting ``Generative AI has just handed them a winning lottery ticket''.

`Paper mills', operated by individuals or businesses, may sell authorship of such articles to researchers, for whom career advancement incentivises participation in such fraud. Paid-for citations, rather than authorship, is an another active area of abuse made more prevalent by AI-generated content. There may also be an unfortunate intersection with the separate advent of paid open-access publication fees which has created an ecosystem of `predatory' journals whose existence is primarily motivated by harvesting such fees rather than executing effective peer review. 

Aside from specifically fraudulent activity, considerable uncertainty has arisen about what is and is not an acceptable level of AI use in preparation of articles and grant submissions. Journals are experiencing a lack of reliable methods to verify or test whether content has been generated via AI, a situation which seems certain to worsen. Two of us have witnessed this directly as journal editors, one of us also experiencing a phenomenon of a journal seeking to automate peer reviewer selection via AI under cover of requests appearing to originate from human editorial board members.

\subsection{Peer review processes}

The established preference for peer-review across academic endeavours generates a considerable workload of unpaid or poorly-paid `service' activity seeking to validate and maintain quality. Generative AI offers strong temptations as a way to alleviate this load.

In reviews of funding applications, the US National Science foundation (NSF) has given explicit notice
that reviewers are forbidden from uploading content to non-approved generative AI tools, citing confidentiality concerns: \href{https://new.nsf.gov/news/notice-to-the-research-community-on-ai?utm_medium=email&utm_source=govdelivery}{NSF Notice to research community}.\footnote{They do not question whether the AI is competent to assist or even carry out a review, but in any case are apparently yet to approve any AI tools.} Notably, proposers of research projects are ``encouraged to indicate in the project description the extent to which, if any, generative AI technology was used and how it was used to develop their proposal'', but are not required to do so. 

There is quite a bit of anecdotal evidence of researchers handing their peer-review duties, particularly of journal articles, over to AI models, despite many journal publishers stating that this is forbidden (again typically citing confidentiality concerns). Consider for instance this editorial in {\bf Nature Biomedical Engineering} \cite{NatureEd} for its vision of a possible future of AI-assisted peer-review. As with AI contributions to article content itself, verification of the presence of this practice ranges from difficult to impossible.

\subsection{Traditional academic methodology}

Research, particularly in the AI domain itself, is increasingly being conducted outside of academia where developers are not necessarily motivated to submit research for peer review and may even be prevented from doing so by commercial considerations, including patenting and shareholder value protection. 

The pace of the technology developments also does not allow the traditional peer-review process to keep up with the rate of research. 

While academics work on drafting manuscripts to report on the state of problem space, the drafts become obsolete before the traditional process of peer review can be concluded. As an example, despite having attempted to include a time stamp for content included in this report, we have had to continuously remove and/or update content that has, in the space of two to three weeks, become obsolete. Attribution of DOIs to online videotaped content is an option to address the rapidly-evolving technology under study.

\section{AI industry energy consumption}\label{Sec:energy-consumption-emmissions}


Artificial intelligence, alongside crypto-mining, is an important driver of a rapid increase in power demand from large datacenters, of which the US alone has almost three thousand, estimated at consuming 5\% of the entire nation's electricity generation (as discussed for example in \href{https://www.washingtonpost.com/business/2024/03/07/ai-data-centers-power/}{The Washington Post}, \href{https://www.bloomberg.com/news/articles/2024-03-26/ai-will-suck-up-500-more-power-in-uk-in-10-years-grid-ceo-says?utm_source=twitter&cmpid%3D=socialflow-twitter-markets&utm_campaign=socialflow-organic&utm_medium=social&utm_content=markets}{Bloomberg}, and \href{https://www.theguardian.com/technology/article/2024/jul/02/google-ai-emissions}{The Guardian}). This is placing considerable strain on utilities, whose capacity-planning horizons are traditionally decades rather than years or months. Moreover, two-thirds of the energy the US produces is `rejected energy,' which does no useful work, and is therefore lost to heat and friction during energy transmission and distribution. Waste heat production from datacenters now exceeds that of the entire aviation industry \cite{neXt} 
and makes a significant adverse contribution to carbon emissions.

Model training and operation are power-intensive processes, with the demand for high-efficiency GPU processors driving the main supplier Nvidia into the world's top-valued companies. Future initiatives to secure provenance and authenticity via blockchain encoding may further heighten demands. It might appear overstated, but the demand for power will increase everywhere. Depending on the country's policy around energy, some countries might have energy issues and vulnerabilities through inequalities in access to power.

\section{Artificial General Intelligence: {\it `The apocalypse'}}\label{Sec:AGI}


Artificial General Intelligence (AGI), the possession by machines of broad capabilities matching or exceeding human across a wide domain of activities, is a topic of academic and intellectual debate (in philosophy, neurosciences, etc.) Some of us, as physicists, have expressed a perspective on why we believe AGI does not pose an immediate risk \cite{marinaremy}.  Long-term thinking about AGI apocalypses might be preventing the addressing of the real dangers associated with non-AGI versions of the technology.

The instigators of the new \href{https://arcprize.org/leaderboard}{Arc million dollar prize}, with challenging tasks for presumptive AGI technologies, claim that progress towards AGI has stalled. In a counterview Yoshua Bengio's \href{https://yoshuabengio.org/2024/07/09/reasoning-through-arguments-against-taking-ai-safety-seriously/?fbclid=IwZXh0bgNhZW0CMTAAAR2dm4S5tVTiVi_XmNH6TXxNQ4MbKkay3lsaqhoC1-0fjP92If_KnUQad74_aem_LnzlpOWyJL_OwIJFiB8UAQ}{blog post} criticises various arguments against the plausibility of AGI being attained in the near future. However even if he is right that these arguments are weak, that does not imply that the opposite will happen since it may just mean that the best arguments have not been considered, or that arguments which appear weak are actually valid.

Outside of considerations of what constitutes AI or machine intelligence, there is disagreement about the meaning of the word `intelligence' itself, e.g.\ Ref.~\cite{JAGI}.
%

%
%
%
%
\section{Doom thinking - AI}\label{Sec:AI-doom}

Recall how pundits responded to the disruptive technology of radio. In the late 1920s, they thought radio would bring world peace, because world leaders would use radio to talk and never have misunderstandings. Then in the 1930s pundits thought radio meant the end of democracies as the ultimate dictators' tool for propaganda. Then we can fast forward around 100 years and radio seems to have not resulted in either the extreme hype nor doom. AI-doom thinking may be repeating the pattern of radio. 

So the need for pragmatism is very high. We should not be surprised if there is dual-use; that said, we should do what we can to make sure conversations are grounded in reality while thinking of 3rd- and 4th-effects, both intended and unintended.

In short, amid the backdrop of different technological revolutions happening, there is no shortage of perceived problems with the most extreme voices capturing the largest amount of airtime and attention. A substantial risk of AI-doom thinking involves amplifying learned helplessness, where individuals feel that there is nothing to be done that will work given the sheer challenges of a topic. At the same time doom thinking might lead to cognitive easing, where repetition of these doom-focused narratives makes citizens more likely to believe them, even if the underlying assumptions and risks warrant closer examination. These will lead to limited, ineffective solutions to these potential problems, see Ref.~\cite{braydoom}.

\section{Addressing challenges on individual level}

\paragraph{Human traits}
As societies we risk misunderstanding the technology as an oracle or sayer of truths. The assignment of human characteristics to a group of software models has historically been documented as potentially leading to nefarious consequences for the organisation of societies \cite{JaronUCB}. On a positive note, and as we discussed in Section~\ref{Sec:energy-consumption-emmissions}, securing enough computing capability and enough power sources to deliver the promises of the GenAI industry might prove to be an important limit on the development and growth of the technology. This is a positive aspect as it may slow down the roll-out of models, giving more time for risk assessment. 
\paragraph{Intentionality} 
Intentionality is one of the best tools to minimise risk of the technology. Being intentional with each of our actions. Maintaining awareness every time we approach a screen or internet device because we are aware that the techniques of persuasion for attracting our attention, often away from our desired goal, are becoming ever more specialised.

\paragraph{Social interactions} 
As communities, we can also choose to spend time together, enjoying each other's company without technology being present, either TVs or background music. Perhaps this may develop better story telling to one another in the family context. Some of us are reminded of times, when traveling in the desert with indigenous peoples: in the desert, at night time, sitting around the camp fire, there is nothing else other than telling stories to entertain each other. 
\paragraph{Time} In uncertain times, time itself is the most valuable commodity. Our own time is the biggest gift we can offer one another. Choosing to invest our energy and time with those we cherish, together and studying content or topics we believe will lead to a better society, and a better tomorrow. This series of White Paper articles, we hope, is an example of one such gift.
%
%
\section{Discussion}

%
\paragraph{Value of volunteer assessment groups.} 
In the space of regulation and AI safety, multiple approaches and multidisciplinarity will be always be needed. In Ref.~\cite{natacad} Eric Horvitz (Microsoft) argues that the work of service of academics and volunteer groups reviewing AI safety in critical systems is severely underfunded. 
\paragraph{Human--AI ecosystem.} He stresses that we are devising risk strategies based on mental models of the world, while 
in reality we are {\it `sticking a toe in the complexity of gigantic social systems where cognitive psychology and large scale interactions --- that we can't predict --- are going to play a huge role'}.\footnote{\href{https://vimeo.com/showcase/11110544/video/936224823/embed}{\it Looking Forward: Reflections from the Workshop Planning Committee.
}} 
Given the complexity of the ecosystems we will be facing in the near future as AI models become tightly interwoven into societies, the development of practices for AI safeguards needs to bring together  social and cognitive psychology solutions, as well as technical and systems engineering solutions. Such safeguards development practices include doing experiments, at scale, factoring in large-scale unknowns, as the reliance on digitization increases. 
\paragraph{Post-epistemic world.}
How do we avoid entering a post-epistemic world \cite{xxx,yyy,zzz}? 
We must not underplay the component brought by `adversarial AI systems' that learn human persuasion techniques in behaviour psychology. In Refs.~\cite{HorvitzAdv, aaa, bbb}, Horvitz discusses adversarial explanations generated by
models that understand the world and human behaviour. These models can take goals from autocratic rulers, and generate believable stories that combine of disinformation and real live events guided by these systems. These might create synthetic histories, including replacing or repositioning content for creating synthetic pasts. We might see civil liberties that we celebrate today coming at risk as Draconian measures, all-encompassing and hard to circumvent, come into play. 

\section{Conclusions: Societal Organisation
}
Our civilisation may appear robust, but it is actually a carefully orchestrated balance. 
Globally, societies rely on densely-coupled supply chains of manufacturing, trade, finance, employment, food, water, transport, energy, technology, and healthcare, and further on delicately balanced networks of geopolitics, law, and order. Each network and supply chain depends on the others through many feedback loops, as was discussed in Ref.~\cite{newscientist}.
In other words civilisation is an adaptive, complex system --- and such systems are susceptible to catastrophic failure.  Loss of any essential subsystem can cause the whole edifice to collapse. Even small glitches can cascade and run out of control. An early example of this was felt in 2008, when local financial failures cascaded through coupled systems to cause a global financial crisis. We believe that the ever more frequent instances of cascade failures are demonstrating our increased vulnerability, which might be severely amplified once the AI component is factored in.
\subsection{AI components in Critical Infrastructures --- Cascade Failures}
Critical infrastructures are considered essential for the normal functioning of societies and their economies, and therefore they are deserving of special protection due to their strategic importance.

\paragraph{Single point of failure.}
Recent cascade-failure like incidents that have taken place (Crowdstrike, \href{https://tfl.gov.uk/campaign/cyber-security-incident}{Transport of London - Cyber-security incident}) have illustrated the dangers of having too much dependence of critical infrastructures on a single source. This certainly translates to AI model validation. The solution is to not permit the existence of single points of failure in critical systems. Part of the mitigation analysis to assess these risks involves the identification those single points of failure, and then the development of designs to either minimise or eliminate the risk associated to match our definition of acceptable risk-level. 

\paragraph{Over-reliance on cloud computing.}
The strong dependence on cloud services has raised questions on this 
failure. It may not be an AI policy issue but it was a process policy issue.  The Crowdstrike failure was amplified by the relative ease granted by the high security privilege of the update. These may not be new perils, but globally speaking we may be facing an over-dependence on cloud services. The wide scope of the event was because it was in the cloud. If we create a dependency on AI similar to the dependence of critical infrastructures on cloud services it creates a single point of failure which might have widespread results. 

\paragraph{System updates --- High-privilege 
security.}
System updating is one of the main components at risk in a global algorithmic pipeline. Software testing is still important, it is not a modern failure, but we can focus on why it caused such a widespread problem. A system update was in security software with very high privilege level. This makes these updates very difficult to roll back. 

\paragraph{Lessons of deployment of AI components in critical infrastructures.}
We are exposed to safety and security issues that affect multiple organisations, multiple people, multiple locations, and multiple countries (without having to mention it is a nuclear level threat). Using AI in critical infrastructure environment does require us to make sure we have all the safety/security mechanism built around it.\\

One of the main points of this first article of our series is to stress that, in opposition to what is often the focus of news media, we see {\bf no threat to human existence stemming from the emergence of, nor the takeover by, Artificial General Intelligence}. Therefore the scope of the work of the standards working group will not involve risk stemming from Artificial General Intelligence in whatever instantiation. Rather we wish to emphasize that the impact from the insertion of narrow AI models on critical infrastructures that support societal organisation has not yet been reliably evaluated nor controlled in a self-consistent and consequential way. As a result of possible consequences to modern civilisation this is the area that we consider significant and worthy of our attention as a working group developing standards on AI Safeguards.

\section{Author short biographies}\label{Sec:Bios}

 \subsection*{Marina Cort\^es}
 
 Marina Cort\^es obtained a Ph.D.\ in Theoretical Physics at the University of Sussex (2008). She was awarded several independent research fellowships at Lawrence Berkeley National  Laboratory (California), University of Cape Town, South Africa, and the Royal Observatory in Edinburgh, where she won a prestigious Marie Curie Fellowship. She is currently Research Faculty at the University of Lisbon, Portugal. Cort\^es’s work has influenced early universe cosmological inflation \cite{MarinaPRL,MarinaSDSS,MarinaBICEP}, and her work on the origin of the arrows of time \cite{MarinaECS} was awarded first place in the Inaugural \href{http://www.buchaltercosmologyprize.org}{Buchalter Cosmology Prize} in 2014. The award recognised their challenging of the time-symmetric laws, and their introduction of the arrow of time back onto the foundations of theoretical physics. Cort\^es has recently founded the new scientific field of Biocosmology \cite{Biocosmology,Biocosmology2,Biocosmology3,IAI, scifoo}.

She is the Chair of IEEE-SA's p3995  working group: \href{https://standards.ieee.org/ieee/3395/11378/}{\it Standard for the Implementation of Safeguards, Controls, and Preventive Techniques for Artificial Intelligence (AI) Models}.

\subsection*{Andrew R.\ Liddle}

Andrew Liddle is a physicist and cosmologist based at the University of Lisbon, Portugal. He obtained his Ph.D.\ at the Universiyt of Glasgow, and has held faculty positions at Imperial College London, at the University of Sussex where he directed the Astronomy Centre for many years, and at the University of Edinburgh. He has worked on major internaitonal proects including the European Space Agency's {\it Planck} satellite and the ongoing Dark Energy Survey. Author of five books and almost three hundred peer-reviewed articles, he is rated by the annual Stanford University review of scientific citation impact \cite{Stanford} as being in the top 0.05\% of scientists worldwide.

\subsection*{Christos Emmanouilidis}

Christos Emmanouilidis (Senior Member, IEEE) carries experience from Industry (Zenon Automation), Academia (University of Groningen, ATHENA Research and Innovation Centre, Cranfield University) and as aninnovation expert (regional government, EC’s JRC and EIT regarding Smart Specialisation Strategies, Industrial Transitions, and Knowledge and Innovation Communities.  With standardisation contributions at CEN (Maintenance), and ISO (Asset Management), he is a Founding Fellow of the International Society of Engineering Asset Management (ISEAM), and vice-chair of IFAC's TC 5.1 Manufacturing Plant Control, chairing WG Advanced Maintenance Engineering Services and Technology. He has had leading project roles  on Human-Centric AI and Cognitive Systems, Robotics and Automation, and Internet of Things.

\subsection*{Anthony E. Kelly}

Anthony E. Kelly earned his Ph.D.\ in Psychological Studies In Education from Stanford University (1988).  He served as faculty member at Rutgers University (1988-2000) and is Professor at George Mason University (2000-present).  He served as a program officer at the US National Science Foundation (NSF) (1997-2000; 2005-2006); and served as a senior advisor at the NSF (2014-2018). His interests in AI converge with career-long interests in advancing interdisciplinary methodologies, particularly the warrants for scientific claims and ethical practices in the social sciences.

\subsection*{Ken Matusow}
Ken Matusow is a serial entrepreneur who has founded numerous technology companies including Q-Group, Detente Technology, Magic Circle Media, BroadBand Central, DiskCorp, and WhozHereNow. He is currently President of Synergicity, an engineering services organization. He is also a travel writer with numerous stories published, both on-line and in print including 'The Best Travel Writing of 2007' and 'The Best Travel Writing of 2009', published by Travelers' Tales, and the creative writing textbook 'Dreams and Inward Journeys' published by Pearson/Longman.

\subsection*{Ragu Ragunathan}

Ragu Ragunathan is an electronics and telecommunication engineer, computer scientist, cybersecurity professional, and an artificial intelligence researcher with a doctoral degree.
He supports U.S. commercial and public sector organizations, ensuring the security compliance of their information systems with legislative requirements, and strengthening their cybersecurity postures.
His professional experience, in the roles of software engineer, program manager, and information technology executive, includes business process re-engineering, increased automations, and productivity enhancements for the U.S. commercial sectors of physical distribution, retail banking, international developmental banking, health insurance, and mortgage lending.
Ragu’s current areas of Artificial Intelligence research include (a) Managing risks, (b) Adoption challenges, (c) Impact on jobs, and (d) Harnessing AI for a better Cybersecurity.

\subsection*{Jayne M.\ Suess}

Jayne Suess is an Independent CyberSecurity Professional with over 34 years of Information Technology, Internal Audit, Information Security, Information Assurance, Governance Risk and Compliance, and Leadership experience. She has worked in the public and private sector including Big-6 Consulting firms, Fortune 500 companies, and small to medium-sized businesses as an employee/consultant. She has also served as the Chief Information Security Officer for small and medium-sized businesses for over 14 years. Doctoral research included exploring factors impacting the performance of critical security controls in protecting consumer personally identifiable information. She holds the following professional security certifications: CISSP, CISA, CCSP, CRISC, CDPSE, and CISM. She is currently independently consulting, serving on Advisory Boards, participating in IEEE/ISO/ IEC Standards working groups (Zero Trust Security Framework, and Implementation Safeguards, Controls, and Technical Directives for AI models).

\subsection*{George Tambouratzis}

George Tambouratzis received a Diploma in Electrical Engineering from N.T.U.A, Athens, Greece, and a Ph.D. in Neural Networks and Pattern Recognition, from Brunel University. He has held a senior research post at Athena Research Centre, Athens, Greece (ILSP) since March 1999. His research interests include Pattern Recognition, Artificial Neural Networks, Computational Intelligence algorithms and Natural Language Processing. He is a Senior Member of IEEE and a member of the IEEE System, Man \& Cybernetics Society and the IEEE Computational Intelligence Society. He has served as the coordinator and lead investigator in several research projects, mainly projects funded by the European Commission. He has also served as an expert for the European Commission.

\subsection*{Janusz Zalewski}

Janusz Zalewski is a Professor Emeritus of Computer Science and Software Engineering at Florida Gulf Coast University, in Fort Myers, Florida, USA, and currently employed as a University Professor at the State Academy of Applied Sciences in Ciechanow, Poland. He previously held academic positions at Embry-Riddle Aeronautical University and University of Central Florida. In the past, he worked on projects for the Superconducting Super Collider, Lawrence Livermore National Laboratory, FAA, and United States Air Force Academy, as well as consulting for a number of private companies, including Lockheed Martin, Harris, and Boeing. He also had fellowships at NASA and Air Force Research Labs. His research interests include application of AI in real-time embedded and cyberphysical systems, safety and security of complex computer systems and networks, and software engineering education. He served as a Secretary of the IEEE Std 1876-2019 ``Networked Smart Learning Objects for Online Laboratories'' and is a member of the IFAC TC 3.1 on Computers for Control.

\subsection*{David A.\ Bray}

David A.\ Bray is a Distinguished Fellow and Chair of the Accelerator, \href{https://www.stimson.org/project/alfred-lee-loomis-innovation-council/}{Loomis Council} at the non-partisan \href{https://www.stimson.org}{Henry L. Stimson Center}. He also is a \href{https://bens.org/people/dr-david-bray/}{Distinguished Fellow} with the \href{https://bens.org}{Business Executives for National Security}. Bray served as Executive Director for the People-Centered Internet coalition chaired by Internet co-originator Vint Cerf.
He led two successful Bipartisan Commissions, including the 2020-2021 \href{https://www.atlanticcouncil.org/content-series/geotech-commission/exec-summary/}{``Geopolitical Impacts of New Tech \& Data} \cite{bray:geopolitical} endorsed by two U.S.\ Senators and two Representatives. In 2016 David Bray was noted as one of \href{https://www.businessinsider.com/23-americans-who-are-changing-the-world-2016-3#david-bray-federal-communication-commission-6}{``24 Americans Who Are Changing the World’’} by \href{https://www.businessinsider.com}{Business Insider}, and later recognized as a \href{https://www.linkedin.com/posts/hunter-muller-924818_technology-techawards-visionaryleadership-activity-7138263332825780225-I6HL/}{``2023 Industry Genius’’} and a ``Global Executive Who Matters'' by \href{https://hmgstrategy.com/}{HMG Strategy}. He was awarded the Department of Defence’s Joint Civilian Service Commendation Award and a National Intelligence Exceptional Achievement Medal. 

\section*{Acknowledgments}
We thank Cat Allman, Paul Anand, Christy Bahn, Catherine Berger, Eduardo Gil Brand\~ao, 
Howard Deiner, Jeanine DeFalco, Chris DiBona, Mark Elliot, Zann Gill, Janet Ginsburg, Vasco Gil Gomes, Shannon Gray, Leighton Johnson, Ted Kahn, Shel Kaphan, Stuart A. Kauffman, Jaron Lanier, Ben Levitan, Gerardo Lisboa, Tim Palmer, Ben Rolfe, Barnaby Simkin, Mike Simmons, Richard Tong, Roopa Vasan, and Aleksandr Zakharchenko for discussions.

\noindent
\rule{\textwidth}{3pt}

\section{Disclaimers}\label{Sec:disclaimers}

This article solely represents the views of a {\bf set of} authors within the IEEE P3395 Working Group, and is not a consensus document. It does not  represent a position of either IEEE or the IEEE Standards Association.\\
%
%
Specifically this document is NOT AN IEEE STANDARD. Information contained in this Work has been created by, or obtained from, sources believed to be reliable, and reviewed by members of the activity that produced this Work. IEEE and the P3395 Working Group (WG) expressly disclaim all warranties (express, implied, and statutory) related to this Work, including, but not limited to, the warranties of: merchantability; fitness for a particular purpose; non-infringement; quality, accuracy, effectiveness, currency, or completeness of the Work or content within the Work. In addition, IEEE and the P3395 WG disclaim any and all conditions relating to: results; and workmanlike effort. This document is supplied “AS IS” and “WITH ALL FAULTS.”

Although the P3395 WG members who have created this Work believe that the information and guidance given in this Work serve as an enhancement to users, all persons must rely upon their own skill and judgment when making use of it. IN NO EVENT SHALL IEEE SA OR P3395 WG MEMBERS BE LIABLE FOR ANY ERRORS OR OMISSIONS OR DIRECT, INDIRECT, INCIDENTAL, SPECIAL, EXEMPLARY, OR CONSEQUENTIAL DAMAGES (INCLUDING, BUT NOT LIMITED TO: PROCUREMENT OF SUBSTITUTE GOODS OR SERVICES; LOSS OF USE, DATA, OR PROFITS; OR BUSINESS INTERRUPTION) HOWEVER CAUSED AND ON ANY THEORY OF LIABILITY, WHETHER IN CONTRACT, STRICT LIABILITY, OR TORT (INCLUDING NEGLIGENCE OR OTHERWISE) ARISING IN ANY WAY OUT OF THE USE OF THIS WORK, EVEN IF ADVISED OF THE POSSIBILITY OF SUCH DAMAGE AND REGARDLESS OF WHETHER SUCH DAMAGE WAS FORESEEABLE.

Further, information contained in this Work may be protected by intellectual property rights held by third parties or organizations, and the use of this information may require the user to negotiate with any such rights holders in order to legally acquire the rights to do so, and such rights holders may refuse to grant such rights. Attention is also called to the possibility that implementation of any or all of this Work may require use of subject matter covered by patent rights. By publication of this Work, no position is taken by the IEEE with respect to the existence or validity of any patent rights in connection therewith. The IEEE is not responsible for identifying patent rights for which a license may be required, or for conducting inquiries into the legal validity or scope of patents claims. Users are expressly advised that determination of the validity of any patent rights, and the risk of infringement of such rights, is entirely their own responsibility. No commitment to grant licenses under patent rights on a reasonable or non-discriminatory basis has been sought or received from any rights holder.

This Work is published with the understanding that IEEE and the p3395 WG members are supplying information through this Work, not attempting to render engineering or other professional services. If such services are required, the assistance of an appropriate professional should be sought. IEEE is not responsible for the statements and opinions advanced in this Work.
~\\


\begin{thebibliography}{99}

%
%
\bibitem{panel4}
US National Academies of Sciences, Engineering and Medicine, 2024, 
\href{https://www.nationalacademies.org/event/41384_04-2024_evolving-technological-legal-and-social-solutions-to-counter-disinformation-in-social-media-a-workshop}{\it Evolving Technological, Legal and Social Solutions to Counter Disinformation in Social Media - A Workshop}, April 11th, 2024.
\href{https://vimeo.com/showcase/11110544/video/936226597}{\it ``Panel 4: Regulatory and Other Incentives and Disincentives for Behavior Change"}.

\bibitem{YT:Weinstein} E. Weinstein, 2024,
\href{https://youtu.be/p_swB_KS8Hw?si=vcADVvP3docOrpgR}{{\it Why does the modern world make no sense?}}, Interview by Chris Williamson at ``Modern Wisdom'', Feb.\ 19th, 2024.

\bibitem{P3395} \href{https://standards.ieee.org/ieee/3395/11378/}{Standard for the Implementation of Safeguards, Controls, and Preventive Techniques for Artificial Intelligence (AI) Models},  IEEE-SA, chair M. Cort\^es.

\bibitem{RAGU2} The White House, 2023, {\it “Executive Order on the Safe, Secure, and Trustworthy Development and Use of Artificial Intelligence}, Oct.\ 30, 2023.  \href{https://www.whitehouse.gov/briefing-room/presidential-actions/2023/10/30/executive-order-on-the-safe-secure-and-trustworthy-development-and-use-of-artificial-intelligence/#:~:text=(a)%20Artificial%20Intelligence%20must%20be}{Available here}. 

\bibitem{regulation}
J. Zalewski, J.M. Suess, G. Tambouratzis, A. E. Kelly, C. Emmanouilidis, R. Martin, R. Ragunathan, M. Cort\^es, A. R. Liddle, and K. Matusow, 2024, {\it AI regulation: an overview of national and international regulatory initiatives and global standards}, to appear.

\bibitem{techno-watch}
G. Tambouratzis, K. Matusow, M. Cort\^es, A. R. Liddle, C. Emmanouilidis, A. E. Kelly, R. Ragunathan, J. M. Suess G. Tambouratzis, and J. Zalewski, 2024, {\it AI Technology Watch: A selection of key developments, emerging technologies and industry trends},
to appear.  

%
\bibitem{AI-health}
M. Cort\^es, A. R. Liddle, A. E. Kelly, D. A. Bray, K. Matusow, R. Ragunathan, J. M. Suess, G. Tambouratzis, and J. Zalewski, 2024, {\it AI and Health: uses and associated risks in the generative AI domain}, to appear.

\bibitem{AI-mental-health}
M. Cort\^es, A. R. Liddle, A. E. Kelly, and G. Northoff, 2024, {\it AI in mental health: a study on the effects of prolonged exposure to the online environment}, to appear. 

\bibitem{Vaswani} A. Vaswani, N. Shazeer, N. Parmar, J. Uszkoreit, L. Jones, A. N. Gomez, L. Kaiser, and I. Polosukhin, 2017, {\it Attention is all you need}, \href{https://arxiv.org/abs/1706.03762}{arXiv:1706.03762 [cs.CL]}.

\bibitem{Hacker} P. Hacker, A. Engel, and M. Mauer, 2023, {\it Regulating ChatGPT and other Large Generative AI Models}, 2023, FAccT '23: Proceedings of the 2023 ACM Conference on Fairness, Accountability, and Transparency, \href{https://doi.org/10.1145/3593013.3594067}{DOI}.

\bibitem{GBGM} R. Gozalo-Brizuela and E. C. Garrido-Merchan, 2023, {\it ChatGPT is not all you need. A State of the Art Review of large Generative AI models}, \href{https://arxiv.org/abs/2301.04655}{arXiv:2301.04655 [cs.LG]}.

\bibitem{Suleyman} M. Suleyman and M. Bhaskar, 2023,{\it The Coming Wave: Technology, Power, and the Twenty-first Century's Greatest Dilemma}, Crown, ISBN 978-0593593950.

\bibitem{Bengio} Y. Bengio et al., 2024, \href{https://assets.publishing.service.gov.uk/media/6655982fdc15efdddf1a842f/international_scientific_report_on_the_safety_of_advanced_ai_interim_report.pdf}{{\it International Scientific Report on the Safety of Advanced AI (Interim report)}}.

\bibitem{BaiDoo} D. Bai'doo-Anu and L. Ansah, 2023, {\it Education in the Era of Generative Artificial Intelligence (AI): Understanding the Potential Benefits of ChatGPT in Promoting Teaching and Learning}, Journal of AI {\bf 7},  52, \href{https://doi.org/10.61969/jai.1337500}{DOI}.

\bibitem{Kotek} H. Kotek, R. Dockum, and D. Sun, 2023, {\it Gender bias and stereotypes in Large Language Models}, Presented at CI '23: Collective Intelligence Conference, Delft, Netherlands, November 2023, \href{https://doi.org/10.1145/3582269.3615599}{DOI}. 

\bibitem{Lucchi} N. Lucchi, 2023, {\it ChatGPT: A Case Study on Copyright Challenges for Generative Artificial Intelligence Systems.} European Journal of Risk Regulation 1--23, \href{https://doi.org/10.1017/err.2023.59}{DOI}. 














\bibitem{Hayes} C. M. Hayes, 2023, {\it Generative Artificial Intelligence and Copyright: Both Sides of the Black Box}, \href{https://doi.org/10.2139/ssrn.4517799}{DOI SSRN}.


\bibitem{stateofai} Nathan Benaich; Air Street Capital, 2023, \href{https://www.stateof.ai/}{State of AI Report 2023}; Oct.\ 12th, 2023;  163 pp; see summary in \href{https://youtu.be/RCRuiu-3VDU }{AI Explained}.

\bibitem{Graceetal} K. Grace, H. Stewart, J. F. Sandk\"uhler, S. Thomas, B. Weinstein-Raun, and J. Brauner, 2024, {\it Thousands of AI Authors on the future of AI}, \href{https://arxiv.org/abs/2401.02843}{arXiv:2401.02843 [cs.CY]}.

\bibitem{RiskRep} P. Slattery, A. K. Saeri, E. A. C. Grundy, J. Graham, M. Noetel, R. Uuk, J. Dao, S. Pour, S. Casper, and N. Thompson, 2024, {\it A systematic evidence review and common frame of reference for the risks from artificial intelligence}, \href{http://doi.org/10.13140/RG.2.2.28850.00968}{DOI}.

\bibitem{SIOshutdown} C. Newton, Z. Schiffer, June 2024, \href{https://www.platformer.news/stanford-internet-observatory-shutdown-stamos-diresta-sio/}{{\it Stanford Internet Observatory shutdown}}, Platformer News.

\bibitem{natacad} 
US National Academies of Sciences, Engineering and Medicine, 2024, 
\href{https://www.nationalacademies.org/event/41384_04-2024_evolving-technological-legal-and-social-solutions-to-counter-disinformation-in-social-media-a-workshop}{\it Evolving Technological, Legal and Social Solutions to Counter Disinformation in Social Media - A Workshop}, April 11th, 2024. \href{https://vimeo.com/showcase/11110544}{Videos available here}.



\bibitem{AIdilemma} T. Harris and A. Raskin, 2023, \href{https://www.youtube.com/watch?v=xoVJKj8lcNQ&t=6s}{{\it The AI dilemma}} on YouTube.

\bibitem{cerftalk} V. Cerf, 2024, \href{https://youtu.be/XrV3yTMCkNg?si=-TbgocRpJOQmKhGm}{The Internet We Deserve: How the History of the Internet Could Inform the Future of Democracy and AI}, interview by the Burnes Center for Social Change (Jan.\ 2024).

\bibitem{BSRS} H. Adkins, B. Beyer, P. Blankinship, A. Oprea, P. Lewandowski, and A. Stubblefield, 2020, {\it Building Secure \& Reliable Systems}, O'Reilly Media, \href{https://sre.google/books/building-secure-reliable-systems/}{Downloadable here}.


\bibitem{DSA} \href{https://www.eu-digital-services-act.com/#:~:text=Under%20the%20Digital%20Services%20Act,right%20to%20freedom%20of%20expression.}{The Digital Services Act (DSA) - Regulation (EU) 2022/2065}.

\bibitem{tencent} L. Chen, 2023, {\it Influence Empire: Inside the Story of Tencent and China’s Tech Ambition}, Hodder, ISBN 978-1529346893.

\bibitem{AI-bill-rights} Conference: ``The Future of Data, Bio, and Algorithms'', \href{https://www.digitalhollywood.com/ai-bill-of-rights---session-four}{Session four: AI Bill of Rights - Ethics and the Law}, M. Martin (Moderator), C. Matthews (Office of Responsible AI, Microsoft), B. Barre (Partner, Le 16 Law); October 19th, 2023.

\bibitem{cerfMarch1st} 
\href{https://www.millennium-project.org/world-futures-day-2024-march-1/}{\it World Futures Day 2024}; March 1st 2024.

\bibitem{china} 
\href{https://www.cac.gov.cn/2023-07/13/c_1690898327029107.htm}
{Office of Central Cyberspace Affairs Commission}, Cyberspace Administration of China, 2024.

\bibitem{CHINAreg} 
\href{https://www.pwccn.com/en/industries/telecommunications-media-and-technology/publications/interim-measures-for-generative-ai-services-implemented-aug2023.html}{Regulatory and legislation: China’s Interim Measures for the Management of Generative Artificial Intelligence Services officially implemented} July 13th, 2023.


\bibitem{Kosseff1} J. Kosseff, 2023, {\it Liar in a Crowded Theater: Freedom of Speech in a World of Misinformation}, Johns Hopkins University Press, ISBN 978-1421447322.

\bibitem{Kosseff2} J. Kosseff, 2022, {\it The United States of Anonymous: How the First Amendment Shaped Online Speech}, Cornell University Press, ISBN 978-1501762383.

\bibitem{Bray-April11th}
D. Bray, 2024, {\it Why, What, and How Free Societies Must Develop Effective Deterrence of Actors Intending Bad Ends Given Our GenAI Era}, \href{https://nsiteam.com/smaspeakerseries_13march2024/}{NSI Speaker session series at US Department of Defense}.

\bibitem{HorvitzAdv} E. Horvitz, 2022, {\it On the Horizon: Interactive and Compositional Deepfakes}, ICMI '22: Proceedings of the 2022 International Conference on Multimodal Interaction
Pages 653--661, \href{https://dl.acm.org/doi/abs/10.1145/3536221.3558175}{DOI}, \href{https://erichorvitz.com/blue_sky_horizon_ICMI.pdf}{PDF here} and \href{https://vimeo.com/showcase/11110544/video/936224823/embed}{video here
}.
\bibitem{aaa} E. Horvitz, 2022,
\href{https://erichorvitz.com/blue_sky_horizon_ICMI.pdf}{Seminar:On the Horizon: Interactive and Compositional Deepfakes}.

\bibitem{bbb} 
\href{https://venturebeat.com/ai/new-deepfake-threats-loom-says-microsofts-chief-science-officer/}{New deepfake threats loom, says Microsoft’s chief science officer}, Venture Beat, September 2022.

\bibitem{OpenAIbio} OpenAI research document, 2024, \href{https://openai.com/research/building-an-early-warning-system-for-llm-aided-biological-threat-creation}{Building an early warning system for LLM-aided biological threat creation}.

\bibitem{OpenAIsen} \href{https://www.washingtonpost.com/documents/2ea97cb4-34df-4bdd-a100-3572e93fdba1.pdf?itid=lk_inline_manual_4}{US Senators public letter to OpenAI}, July 2024. 

\bibitem{MicrosoftAItransp} \href{https://www.microsoft.com/en-us/corporate-responsibility/responsible-ai-transparency-report}{Microsoft Responsible AI Transparency Report}, May 2024.




\bibitem{Starlink} Business Standard, 2023, \href{https://www.business-standard.com/world-news/spacex-s-starlink-clears-military-tests-in-arctic-paving-way-for-contracts-123120700086_1.html}{SpaceX's Starlink clears military tests in Arctic, paving way for contracts}.







\bibitem{KHillbook} K. Hill, 2023, {\it Your Face Belongs to Us: The Secretive Startup Dismantling Your Privacy}, Random House, ISBN 978-0593448564.

\bibitem{BrayDignity} D. A. Bray, 2024, \href{https://www.linkedin.com/pulse/digital-dignity-information-discernment-advancing-data-bray-phd-sktve}{Digital Dignity and Information Discernment: Advancing People-Centered Approaches to Data and AI} [via LinkedIn].

\bibitem{Merton} R. K. Merton, 1936, {\it The Unanticipated Consequences of Purposive Social Action}, 
American Sociological Review
Vol. 1, No. 6, 894, accessible \href{https://www.jstor.org/stable/2084615}{here in JSTOR}.

\bibitem{CoSAI} \href{https://www.coalitionforsecureai.org/wp-content/uploads/2024/07/CoSAI-Charter-FINAL.pdf}{Coalition for Secure AI charter} (July 2024). 

\bibitem{huggingface} \href{https://huggingface.co/spaces/open-llm-leaderboard/blog}{Hugging Face LLM leaderboard blog.}

\bibitem{harvard-misinformation} \href{https://misinforeview.hks.harvard.edu/wp-content/uploads/2023/07/altay_survey_expert_views_misinfo_20230727.pdf}{A survey of expert views on misinformation: Definitions, determinants, solutions, and future of the field}, Harvard Kennedy School Misinformation Review (2023) Volume 4; Issue 4; \href{https://doi.org/10.37016/mr-2020-119}{DOI}.

\bibitem{Augenstein} I. Augenstein et al., 2023, {\it Factuality Challenges in the Era of Large Language Models}, \href{https://arxiv.org/abs/2310.05189}{arXiv:2310.05189 [cs.CL]}.

\bibitem{GeorgeRef1} G. Tambouratzis, 2023, {\it ChatGPT Multilingual Querying Consistency – A Test Case}, ERCOM News 136, available \href{https://ercim-news.ercim.eu/en136/special/chatgpt-multilingual-querying-consistency---a-test-case}{here}.

\bibitem{GeorgeRef2} S. Shafayat, E. Kim, J. Oh, and A. Oh, 2024, {\it Multi-FAct: Assessing Multilingual LLMs’ Multi-Regional Knowledge using FActScore}, \href{https://arxiv.org/abs/2402.18045}{arXiv:2402.18045 [cs.CL]}.

\bibitem{braycerfmisinformation} D. A. Bray and V. Cerf, 2024, \href{https://www.thecipherbrief.com/column_article/tech-should-advance-standards-to-assess-information-quality}{{\it Tech Should Advance Standards to Assess Information Quality}}  April 15th, 2024.

\bibitem{woolley} S. C. Woolley and P. N. Howard, 2016, {\it Automation, algorithms, and politics| political communication, computational propaganda, and autonomous agents—Introduction}, International Journal of Communication 10, accessible
\href{https://scholar.google.com/citations?view_op=view_citation&hl=en&user=8p_7dIkAAAAJ&citation_for_view=8p_7dIkAAAAJ:YFjsv_pBGBYC}{here}; 
P. N. Howard, B. Kollanyi, and S. Woolley, 2016, {\it Bots and Automation over Twitter during the US Election}, Computational propaganda project: Working paper series 21, Issue 8, accessible \href{https://scholar.google.com/citations?view_op=view_citation&hl=en&user=8p_7dIkAAAAJ&citation_for_view=8p_7dIkAAAAJ:Y0pCki6q_DkC}{here}; 
S. Woolley and D. Guilbeault, 2017, {\it Computational propaganda in the United States of America: Manufacturing consensus online}, Computational Propaganda Project, accessible \href{https://scholar.google.com/citations?view_op=view_citation&hl=en&user=8p_7dIkAAAAJ&citation_for_view=8p_7dIkAAAAJ:LkGwnXOMwfcC}{here}. 

\bibitem{Donovan1} \href{https://www.judiciary.senate.gov/imo/media/doc/Donovan%20Testimony%20(updated).pdf}{Hearing on {\it Algorithms and Amplification: How Social Media Platforms' Design Choices Shape Our Discourse and Our Minds}}, Statement of Joan Donovan, before the Senate Committee on the Judiciary Subcommittee on Privacy, Technology and the Law (April 27th 2021).

\bibitem{Donovan2} J. Donovan, 2020, \href{http://opentranscripts.org/transcript/true-costs-of-misinformation/}{{\it True costs of misinformation}}, presentation at the Berkman Klein Center for Internet \& Society.

\bibitem{Chan-LLMAttacks} S. Shan, A. Bhagoji, H. Zheng, and B. Zhao, 2022, {\it Poison Forensics: Traceback of Data Poisoning Attacks in Neural Networks}, 31st USENIX Security Symposium (USENIX Security 22). \href{https://www.shawnshan.com/files/publication/forensics.pdf}{Available here}. 

\bibitem{persuade1} M. Burtell and T. Woodside, 2023, {\it Artificial Influence: An Analysis Of AI-Driven Persuasion}, \href{https://arxiv.org/abs/2303.08721}{arXiv:2303.08721 [cs.CY]}.

\bibitem{persuade2} F. Salvi, M. H. Ribeiro, R. Gallotti, and R. West, 2024, {\it On the Conversational Persuasiveness of Large Language Models: A Randomized Controlled Trial}, \href{https://arxiv.org/abs/2403.14380}{arXiv:2403.14380 [cs.CY]}.

\bibitem{JaronUCB} J. Lanier, 2023,
\href{https://youtu.be/W2OwzfEaasI}{{\it Data Dignity and the Inversion of AI}}, at UC Berkeley; Sept.\ 2023.

\bibitem{NatureEd} Editorial: {\it The advent of human-assisted peer review by AI}, Nat. Biomed. Eng. {\bf 8}, 665 (2024). \href{https://doi.org/10.1038/s41551-024-01228-0}{DOI}.


\bibitem{neXt} T. Jackson and I. R. Hodgkinson, 2024, \href{https://www.weforum.org/agenda/2022/10/dark-data-is-killing-the-planet-we-need-digital-decarbonisation/?utm_campaign=social_video_2022&utm_content=27617_dark_data&utm_medium=social_video&utm_source=linkedin&utm_term=1_1}{World Economic Forum Climate Action report, 2022}; 
\href{https://gail.world/next-forum-food/}{NeXt Forum}, Global Action Improv Lab; March 27th, 2024
 \href{https://youtu.be/t1hFi4TnkhI}{Session recording available here}.


\bibitem{marinaremy} M. Cort\^es, 2023, {\it Free will and the arrow of time}, in `Time and Science Vol 3: Physical Sciences and Cosmology', eds R. Lestienne and P. Harris, World Scientific, July 2023, 23 pages, ISBN 978-1800613768.

\bibitem{JAGI} Introduction to JAGI special issue ‘On defining artificial intelligence’, Journal of Artificial General Intelligence  {\bf 11}, 1 (2024), \href{https://intapi.sciendo.com/pdf/10.2478/jagi-2020-0003#page=4}{Online here}.

\bibitem{braydoom} D. A. Bray, 2024, \href{https://www.oodaloop.com/archive/2024/04/30/houston-we-have-a-problem-lets-get-to-solving-it/}{\it Houston, We Have a Problem... (Let’s Get to Solving It)}, OODA LLC, April 30th, 2024.

\bibitem{xxx} S. Fuller, 1994, {\it The Sphere of Critical Thinking in a Post Epistemic World}, Informal Logic {\bf 16}, 1, \href{https://doi.org/10.22329/il.v16i1.2434}{DOI}. 

\bibitem{yyy} J. Shaw, 2021, {\it Feyerabend and Manufactured Disagreement: Reflections on Expertise, Consensus, and Science Policy}, Synthese {\bf 198}, 6053,
\href{https://www.academia.edu/37789209/Feyerabend_on_Disagreement_and_Expertise_in_a_Free_Society_Reflections_on_Consensus_and_Science_Policy}{available here}.

\bibitem{zzz} I. J. Kidd, 2016, {\it What’s So Great About Science?” Feyerabend on the Ideological Use and Abuse of Science}, in E. Aronova and S. Turchetti (eds.), {\it The Politics of Science Studies}. pp. 55-76, \href{https://philarchive.org/rec/KIDWSG}{available here}.


\bibitem{newscientist} D. MacKenzie, 2016, {\it What if civilisation collapses?}, New Scientist {\bf 3100}, 45.

\bibitem{MarinaPRL} A. R. Liddle and M. Cort\^es, 2013, {\it Cosmic microwave background anomalies in an open universe}, 
Phys. Rev. Lett. {\bf 111}, 111302 [\href{https://arxiv.org/abs/1306.5698}{arXiv:1306.5698 [astro-ph.CO]}].

\bibitem{MarinaSDSS} SDSS Collaboration. H. Aihara et al.\ inc.\ M. Cort\^es, 2011, {\it The Eighth Data Release of the Sloan Digital Sky Survey: First Data from SDSS-III}, Astrophys. J. Supp. {\bf 193}, 29 [\href{https://arxiv.org/abs/1101.1559}{arXiv:1101.1559 [astro-ph.IM]}].

\bibitem{MarinaBICEP} M. Cort\^es, A. R. Liddle, and D. Parkinson, 2015 {\it Tensors, BICEP2, prior dependence, and dust}, Phys. Rev. D {\bf 92}, 063511 [\href{https://arxiv.org/abs/1409.6530}{arXiv:1409.6530 [astro-ph.CO]}].

\bibitem{MarinaECS} M. Cort\^es and L. Smolin, 2014, {\it The Universe as a Process of Unique Events}, Phys. Rev. D {\bf 90}, 084007 [\href{https://arxiv.org/abs/1307.6167}{arXiv:1307.6167 [gr-qc]}].

\bibitem{buchalter}\href{http://www.buchaltercosmologyprize.org}{Buchalter Cosmology Prize}, 2014, M. Cort\^es and L. Smolin, First Place, USD 10k.

\bibitem{Biocosmology} M. Cort\^es, S. A. Kauffman, A. R. Liddle, and L. Smolin, 2022a, {\it Biocosmology: Towards the birth of a new science}, \href{https://arXiv.org/abs/2204.09378}{arXiv:2204.09378}.

\bibitem{Biocosmology2} M. Cort\^es, S. A. Kauffman, A. R. Liddle, and L. Smolin, 2022b, {\it Biocosmology: Biology from a cosmological perspective}, \href{https://arXiv.org/abs/2204.09379}{arXiv::2204.09379} 

\bibitem{Biocosmology3} M. Cort\^es, S. A. Kauffman, A. R. Liddle, and L. Smolin, 2022c, {\it The TAP equation: evaluating combinatorial innovation in Biocosmology}, \href{https://arXiv.org/abs/2204.14115}{arXiv:2204.14115}.

\bibitem{IAI} M. Cort\^es, 2022, \href{https://iai.tv/articles/the-most-complex-thing-in-the-universe-auid-2110}{\it The most complex thing in the universe.
Biocosmology: the birth of a new science?}, published by the Institute of Arts and Ideas; 

\bibitem{scifoo} M. Cort\^es, May 2021, {\it Biocosmology: birth of a new science}, presented at SciFoo, Google, Mountainview, CA, US, \href{https://www.youtube.com/watch?v=9TH4UsyE3fo}{Video}.

\bibitem{Stanford} J. P. A. Ioannidis, 2023, {\it Updated science-wide author databases of standardized citation indicators},
\href{https:doi.org/10.17632/btchxktzyw.6}{DOI}.

\bibitem{bray:geopolitical} D. A. Bray, P. Brooks, S. Wander, J. Goodman, and T. Carlson, 2021, \href{https://papers.ssrn.com/sol3/papers.cfm?abstract_id=4075200}{\it Report of the Commission on the Geopolitical Impacts of New Technologies and Data}, Commission on the Geopolitical Impacts of New Technologies and Data.


%














































\end{thebibliography}
\end{document}